\documentclass[twocolumn]{aastex631}

\newcommand{\angstrom}{\textup{\AA}}

\newcolumntype{b}{X}
\newcolumntype{s}{>{\hsize=.5\hsize}X}

\usepackage{textcomp}
\usepackage{CJKutf8}
\usepackage{amsmath}
\usepackage{color,soul}
\usepackage{tabularx}

\begin{document}

\title{Measuring The Mass-Radius Relation of White Dwarfs Using Wide Binaries}

\author[0000-0002-6270-8624]{Stefan Arseneau}
\affiliation{Department of Physics and Astronomy, Johns Hopkins University, Baltimore, MD 21218, USA}

\author[0000-0002-0572-8012]{Vedant Chandra}
\affiliation{Center for Astrophysics $\mid$ Harvard \& Smithsonian, 60 Garden St, Cambridge, MA 02138, USA}

\author[0000-0003-4250-4437]{Hsiang-Chih Hwang}
\affiliation{Institute for Advanced Study $\mid$ 1 Einstein Drive, Princeton, NJ 08540, USA}

\author[0000-0001-6100-6869]{Nadia L. Zakamska}
\affiliation{Department of Physics and Astronomy, Johns Hopkins University, Baltimore, MD 21218, USA}

\author[0000-0002-5864-1332]{Gautham Adamane Pallathadka}
\affiliation{Department of Physics and Astronomy, Johns Hopkins University, Baltimore, MD 21218, USA}

\author[0000-0002-8866-4797]{Nicole R. Crumpler}
\affiliation{Department of Physics and Astronomy, Johns Hopkins University, Baltimore, MD 21218, USA}

\author[0000-0001-5941-2286]{J.J. Hermes}
\affiliation{Department of Astronomy \& Institute for Astrophysical Research, Boston University, 725 Commonwealth Ave., Boston, MA 02215, USA}

\author[0000-0002-6871-1752]{Kareem El-Badry}
\affiliation{Department of Astronomy, California Institute of Technology, 1200 East California Boulevard, Pasadena, CA 91125, USA}

\author[0000-0003-4996-9069]{Hans-Walter Rix}
\affiliation{Max-Planck-Institut f\"uur Astronomie, K\"onigstuhl 17, D-69117 Heidelberg, Germany}

\author[0000-0002-3481-9052]{Keivan G.\ Stassun}
\affiliation{Department of Physics and Astronomy, Vanderbilt University, Nashville, TN 37235, USA}

\author[0000-0002-2761-3005]{Boris T. G\"ansicke}
\affiliation{Department of Physics, University of Warwick, Coventry CV4 7AL, UK}

\author[0000-0002-8725-1069]{Joel R. Brownstein}
\affiliation{Department of Physics and Astronomy, University of Utah, 115 S. 1400 E., Salt Lake City, UT 84112, USA}

\author{Sean Morrison}
\affiliation{Department of Astronomy, University of Illinois at Urbana-Champaign, Urbana, IL 61801, USA}

\begin{abstract}

\noindent
Measuring the mass-radius relation of individual white dwarfs is an empirically challenging task that has been performed for only a few dozen stars. 
We measure the white dwarf mass-radius relation using gravitational redshifts and radii of 137 white dwarfs in wide binaries with main sequence companions. 
We obtain the space velocities to these systems using the main sequence companion, and subtract these Doppler redshifts from the white dwarfs' apparent motions, isolating their gravitational redshifts. 
We use \textit{Gaia} data to calculate the surface temperatures and radii of these white dwarfs, thereby deriving an empirical gravitational redshift-radius relation. 
This work demonstrates the utility of low-resolution Galactic surveys to measure the white dwarf equation of state. 
Our results are consistent with theoretical models, and represent the largest sample of individual white dwarf gravitational redshift measurements to date.

\end{abstract}

\section{Introduction} \label{sec:intro}

White dwarfs represent the end stage in evolution of the majority of stars in the Galaxy. These are stars that have exhausted their nuclear fuel and shed their outer layers, revealing a degenerate core. Their high density (with masses comparable to that of the Sun and radii on the order of $0.01$ \(R_\odot\)) enables observations of quantum mechanical phenomena occurring at macroscopic scales. After formation, with no way to generate energy through nuclear fusion, white dwarfs gradually cool, with observed surface temperatures decreasing from $>100,000$ K to $<3,000$ K. 

Since white dwarfs do not undergo nuclear fusion, their radii are determined by the balance of gravitational forces acting to compress the star and electron degeneracy pressure acting against collapse. As a consequence, white dwarfs with larger masses have smaller radii \citep{10.1093/mnras/93.5.390}. This relation is governed by the equation of state (EoS) of the star. Directly measuring the equation of state of white dwarfs is an open problem in astrophysics with wide-ranging implications \citep{tremblay_dr1}. High-mass white dwarfs are hypothesized to be progenitors of Type Ia supernovae, which are used to measure the accelerating expansion of the Universe  \citep{reiss_1998, Perlmutter_1997}. Understanding the EoS is crucial to understanding mechanisms that could generate Type Ia supernovae and the explosion process \citep{hillebrandt_2000}. Additionally, white dwarfs can be used as cosmic clocks in Galactic archaeology \citep{wood_1992, Fontaine_2001, hansen_2002, Cukanovaite_2023}. Precise modeling of their cooling allows a measurement of the age of stellar populations, and understanding the EoS is crucial for this analysis \citep{oswalt_1996}.

While to the leading order white dwarfs cool at an almost constant radius determined by their mass and the EoS, there are interesting small deviations which depend on the surface temperature and the core composition. The majority of the mass of the white dwarf is contained in the degenerate isothermal core, with a small fraction of mass forming the non-degenerate opaque photosphere \citep{Fontaine_2001} which determines the white dwarf's appearance to the observer.  Because of the contribution of the thermal pressure in hotter white dwarfs, they have greater radii than cooler stars of the same mass. Likewise, at a fixed radius a hotter white dwarf should have a greater mass, to fit into the same radius with a puffier atmosphere.

Due to their high surface gravity, white dwarfs are believed to be highly stratified. At the top of their atmosphere is a thin layer of hydrogen, and beneath that is a helium layer. Their core composition is not known exactly, but is believed to be dominated by either oxygen and neon, or carbon and oxygen depending on the mass of the star. Small differences of the nucleon-to-electron ratio in these core compositions can be probed by a precise measurement of the mass-radius relation. In the high-mass limit, models with O/Ne core compositions are predicted to be more compact than their C/O counterparts \citep{refId0}.

The surface temperature $T_{\text{eff}}$ and radius $R$ of a white dwarf can be measured by fitting model atmospheres to observed photometry, as long as the star's distance from the observer is known. In order to test the mass-radius relation, this photometric measurement of the radius must be supplemented by the measurement of the star's mass $M$. \cite{https://doi.org/10.1002/andp.19163540702} predicted that photons propagating from a gravitational potential well (such as the surface of a white dwarf) should shift to redder wavelength as they leave. This results in a measurable redshift of the observed wavelengths of absorption lines in the white dwarf's spectrum from their laboratory wavelengths by an amount $\propto M/R$, which has been observed previously \citep{greenstein_gr, greenstein_gr2, koester_gr, bergeron_gr, reid_gr}. Therefore, in principle the combination of the photometric radius and the gravitational redshift fully constrains the EoS. However, in practice, there is a degeneracy between this gravitational redshift and the Doppler shift due to the star's physical radial velocity relative to the observer.

Several methods exist to disentangle the gravitational redshift from the radial velocity. \cite{Chandra_2020} measured the white dwarf mass-radius relation statistically, by averaging out the radial velocities of 3,308 white dwarfs, binning the population as a function of radius. Measurements of the mass-radius relation using redshifts in individual stars have been made by \cite{Pasquini_2019}. They broke the degeneracy by observing six white dwarfs in the Hyades cluster, exploiting the fact that they were in a comoving cluster with other stars whose radial velocity is known. \cite{parsons_2017} measured the mass-radius relation of 16 white dwarfs in eclipsing binaries with low-mass main sequence stars where the Doppler shifts can be calculated from the orbital dynamics. 

Binaries of white dwarfs with main sequence stars provide an opportunity to disentangle Doppler shifts and gravitational redshifts because the gravitational redshift in main sequence stars -- which are much larger than white dwarfs at a similar mass -- is negligible in comparison to that of white dwarfs \citep{elbadry2022gravitational}. For example, \cite{10.1093/mnras/sty2404} measured the gravitational redshift of Sirius B by computing the difference in the apparent radial velocity between it and Sirius A -- its main sequence companion. In combination with the radius known from photometric measurements, the gravitational redshift measurement yields mass and therefore a constraint on the EoS. A similar technique was used by \citet{greenstein}, \citet{Provencal_1998}, and \citet{Bond_2017} on several white dwarfs in binaries with well-characterized orbital parameters. In these systems, masses inferred from gravitational redshifts can be compared with the dynamical masses from the orbital solutions. 

In this work, we explore the ability of modern large-scale Galactic surveys to probe the mass-radius relationship of white dwarfs and to test its temperature and chemical composition dependence. To this end, we measure the gravitational redshifts of 137 individual white dwarfs in wide binaries with main sequence stars. We compare the apparent radial velocities of the main sequence and white dwarf companions, leveraging the small orbital motions of wide binaries, to break the degeneracy between radial velocity and gravitational redshift. This, in combination with other data enables us to measure their mass-radius relation. We describe our sample in Section \ref{sec:data}. In Section \ref{sec:methods}, we describe our radius and radial velocity measurements, as well as our process for cleaning the sample. We present our main results, gravitational redshift as a function of radius and the effect of temperature in the sample, in Section \ref{sec:results}. Finally, we discuss our findings in Section \ref{sec:discussion}. Surface gravity $\log g$ is reported as the base-ten logarithm of the value in cgs units and all wavelengths are as measured in vacuum.  


\section{Data and sample selection} \label{sec:data}

\subsection{SDSS Spectroscopy}

\begin{table}[]
    \centering
    \begin{tabularx}{\columnwidth}{bs}
        \hline
        Measurements & \# of Systems  \\
        \hline
        Reliable WDMS Wide Binary & 14,685 \\
        Has MS RV & 611 \\
        Has DA WD RV & 305 \\ 
        \hline
        \textbf{Total after Cleaning} & \textbf{137}
    \end{tabularx}
    \caption{Sample size of suitable WDMS binaries with DA white dwarfs at each step of sample selection. 4 WD RVs from SDSS I-IV come from the Apache Point Observatory. The sample cleaning step is described in detail in Section \ref{sec:cleaning}. The makeup of the final sample is presented in Table \ref{tab:final_sample}.}
    \label{tab:raw_sample}
\end{table}

This paper uses data from the Sloan Digital Sky Survey (SDSS). The currently ongoing fifth generation of the SDSS (SDSS-V; \citealt{sdss-v}) started operations in November 2020 using the Apache Point Observatory 2.5 m telescope \citep{gunn_2006} and later expanded to use the Las Campanas Observatory 2.5 m telescope \citep{bowen_1973}. One of the constituent projects, the Milky Way Mapper, aims to obtain multi-epoch spectroscopy of several million stars in the Galaxy using the Baryon Oscillation Spectroscopic Survey spectrograph (BOSS; \citealt{Smee_2013}) and the Apache Point Observatory Galactic Evolution Experiment spectrographs (APOGEE; \citealt{wilson_2019}). All the SDSS-V data used in this paper are from the BOSS spectrograph, which is a low-resolution spectrograph with range $3,650\,\angstrom -9,500\,\angstrom$ at resolution $R\sim 1,800$. The absolute wavelength calibration of BOSS data in SDSS-V is at the design limit of the spectrograph, accurate at better than $7$ km s$^{-1}$ although for white dwarfs, which tend to be faint, the signal-to-noise ratio of the spectrum often is the limiting factor for radial velocity measurements. The BOSS data reductions used in this paper were obtained using the reduction pipeline \verb|v6_1_0|.

The previous generations of the SDSS used an earlier spectrograph with the wavelength coverage of $3,800\,\angstrom-9,000\,\angstrom$ at a resolution of $R\sim 1,800$ \citep{york_2000} until the upgrade to BOSS. The accuracy of the absolute wavelength calibration of SDSS I-IV data is $7$ km s$^{-1}$. All SDSS I-IV data are public. We use both the final data release of SDSS I-IV as well as the data from SDSS-V as of 2023-08-01.

\subsection{Sample Selection} \label{subsec:sample}

Our parent sample is the catalog of 22,563 wide binaries consisting of one main sequence star and one white dwarf from \cite{El_Badry_2021}. Wide binaries in this catalog were selected on the basis of the two companions having a similar parallax and a similar proper motion. For each system in this sample, we first identify the white dwarf and the main-sequence star by their positions on the color-magnitude diagram. Chance alignments are removed from the sample by requiring that {\tt\string R\_chance\_align} $\leq 0.01$, as recommended by \cite{El_Badry_2021}. This value is calculated using the astrometric parameters of both potential objects in the binary and represents a pseudo-probability of chance alignments. 14,685 wide binaries remain after this step. 

We then cross-match the main-sequence companions against the set of \textit{Gaia} DR3 objects with radial velocities measured to an uncertainty less than $7$~km s$^{-1}$. We choose this as the largest uncertainty that, when combined in quadrature with a typical white dwarf radial velocity uncertainty, yields gravitational redshift uncertainties of approximately $15$ km s$^{-1}$. This is the accuracy necessary to detect the gravitational redshift of a typical $0.6$ M$_\odot$ white dwarf at two standard deviations. \textit{Gaia} radial velocities are calculated using the high-resolution radial velocity spectrometer ($R \sim 11,500$). The \textit{Gaia} pipeline estimates atmospheric parameters for each star and then calculates radial velocities by $\chi^2$ minimization on a grid of $-1,000$~km s$^{-1}$ to $1,000$~km s$^{-1}$ \citep{Katz_2023}. 

There are 470 binaries in the \citet{El_Badry_2021} catalog for which the main sequence companion has a \textit{Gaia} radial velocity with the requisite uncertainty, and their median brightness is $G \approx 13.4$~mag. Stars that do not have radial velocities in \textit{Gaia} are cross-matched against the spectroscopic database of the SDSS, so that their radial velocities can be measured from their spectra. There are 137 additional binaries in the \citet{El_Badry_2021} catalog for which the main sequence companion has a spectrum in SDSS. Next, we cross-match the white dwarf companions from \citet{El_Badry_2021} against the SDSS spectroscopic database, obtaining 603 matches. We visually inspect all spectra and remove those that do not have a hydrogen-dominated atmosphere (DA).  

\begin{figure}
    \centering
    \includegraphics[width=\columnwidth]{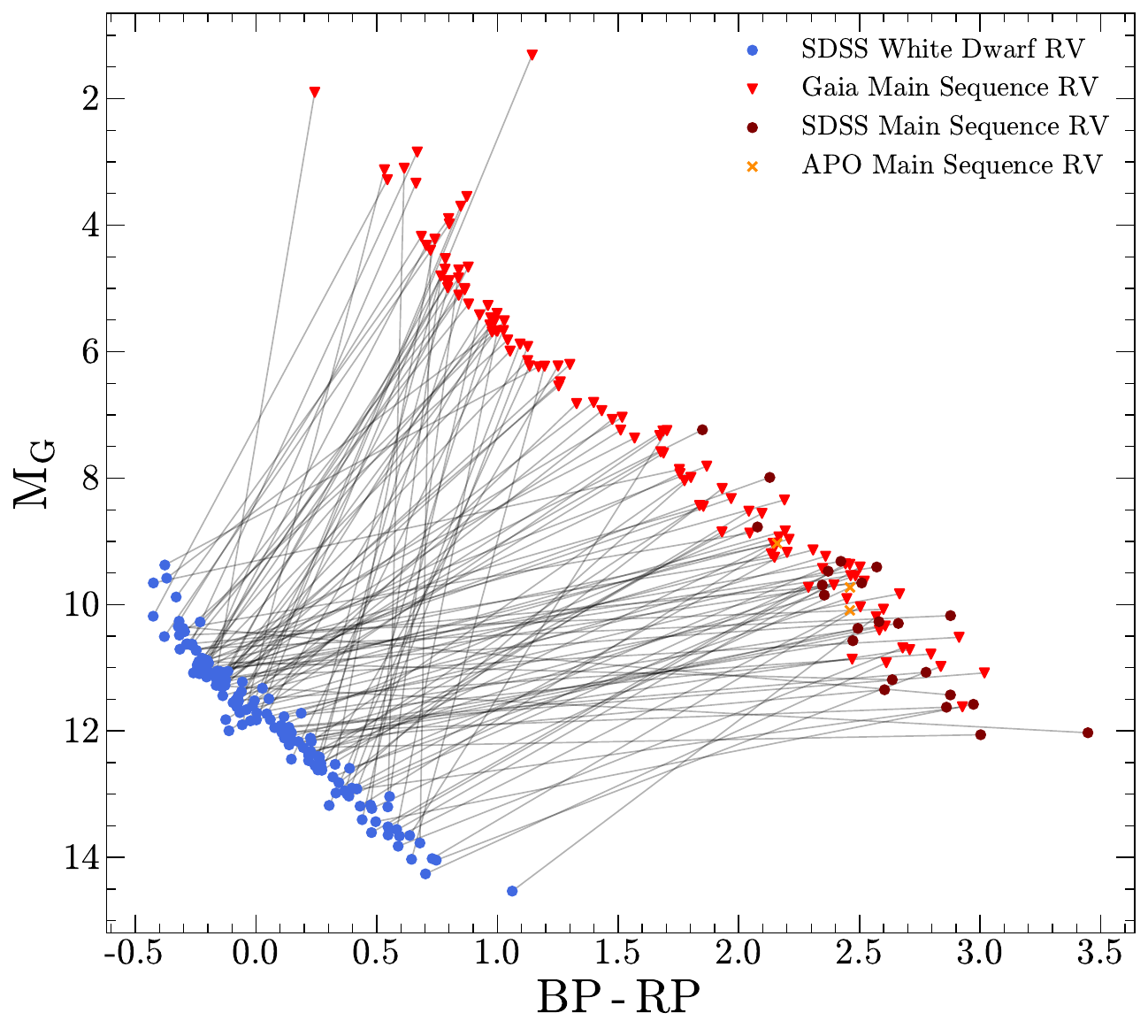}
    \caption{Color-magnitude diagram of the WD+MS wide binary sample. Orbiting companions are connected by gray lines.}
    \label{fig:cmd}
\end{figure}

This leaves us with 295 systems where spectra of both components are available either from SDSS or from {\it Gaia}. Of these, 125 have white dwarf spectra from SDSS-V and 170 come from SDSS I-IV. By spectrograph, 79 white dwarf spectra were observed using the original SDSS spectrograph from the first and second generations of SDSS and 216 with BOSS. An additional 6 white dwarf radial velocities come from the high-resolution catalog of \cite{Falcon_2010}. Most main sequence stars in the sample, 214, have radial velocities measured to high precision in \textit{Gaia}. In addition we obtain 75 main-sequence spectra from SDSS-V and 6 from SDSS I-IV, all of them from BOSS. Figure \ref{fig:cmd} presents the color-magnitude diagram of the sample, and Table \ref{tab:raw_sample} presents the size of the sample at each stage of the sample selection. 

\subsection{Apache Point Observations}
We observed four main sequence stars in binary systems with white dwarfs using the 3.5 m telescope at the Apache Point Observatory in Sunspot, New Mexico. These objects were selected such that their white dwarf companions had spectra in SDSS-IV. The observations were carried out on 2020-02-16 (PI: Hsiang-Chih Hwang) using the Dual Imaging Spectrograph (DIS) with the R1200 grating and a $1.5$~arcsec slit. The wavelength coverage of this filter is $6,000\angstrom - 7,000\angstrom$. This allows observations with resolution $R\sim 4,000$. Each target was observed in two exposures of $300$ seconds. The data are reduced using the standard IRAF pipeline.

The stars were originally identified as being in comoving wide binaries in \textit{Gaia} DR2 astrometry \citep{Hwang2020, Hwang2021}. All four targets were later included in the wide binary catalog by \cite{El_Badry_2021}. 




\section{Methods} \label{sec:methods}

\subsection{Gravitational Redshift Measurement}

General relativity predicts that photons acquire an apparent redshift as they propagate from the white dwarf's gravitational potential to the observer at infinity, equal to 
\begin{equation}
    v_g = \frac{\delta\lambda}{\lambda_0}c = \frac{GM}{cR},
\end{equation}
where $\lambda_0$ is the photon's laboratory-frame wavelength. For a fully degenerate non-relativistic equation of state appropriate for mid-to-low mass white dwarfs, their mass and radius are related via 
\begin{equation}
    M \propto R^{-3}.
\end{equation}
For such objects, we expect to see
\begin{equation}
    v_g \approx 10^{-6} \text{ km s}^{-1} \left(\frac{R}{\text{R}_\odot}\right)^{-4}.
\end{equation}
For a typical white dwarf with mass $0.6$ M$_{\odot}$, the radius is $0.014$ R$_\odot$, and therefore the gravitational redshift is $30$ km s$^{-1}$, detectable in Galactic radial velocity surveys such as SDSS. In contrast, for a main-sequence Sun-like star the gravitational redshift is $v_g=0.6$ km s$^{-1}$, negligibly small compared to the typical gravitational redshift for white dwarfs.

Although radius is the dominant parameter in determining the mass of a white dwarf, smaller effects on the mass-radius relation exist due to temperature, core composition, and hydrogen and helium mass. These variations are calculated in \cite{Fontaine_2001}. Because $v_g$ is dependent on both mass and radius, it provides one of two independent constraints needed to measure the mass-radius relation of white dwarfs. 

\subsection{\texttt{corv}: White Dwarf Radial Velocities}

The apparent radial velocities of white dwarfs are composed of the gravitational redshift and the Doppler effect associated with their motion through the Galaxy. The typical approach for calculating radial velocities using spectral absorption lines is to first identify and isolate the wavelength ranges of interest surrounding the lines. Then, a Gaussian or a Voigt profile is fitted to the isolated region around the spectral line, and the resulting line centroids $\lambda_{\text{obs}}$ can be converted to apparent radial velocities by 
\begin{equation}
    v_{\text{app}} = \left(\frac{\lambda_{\text{obs}} - \lambda_0}{\lambda_0}\right)\cdot c.
\end{equation}
However, without additional care the Stark broadening of absorption lines in high $\log g$ stars and the relatively low resolution of SDSS spectra result in low-quality measurements of radial velocity.


We calculate the white dwarf radial velocities from their spectra using the Compact Objects Radial Velocities (\texttt{corv}\footnote{\url{https://github.com/vedantchandra/corv}}) python library which is specifically designed to measure white dwarf radial velocities. \texttt{corv} gains statistical power by fitting several lines at once, either by fitting Voigt profiles with the same centroid to multiple lines or by cross-correlating the spectrum with a theoretical atmospheric model. In the latter method, \texttt{corv} can fit a DA model atmosphere from \cite{Koester2010} to the spectrum in windows around the Balmer absorption lines \citep{chandra_2020_wdtools} and then obtain the precise radial velocity via cross-correlation. We use the latter method which provides a more realistic estimator of the shape of the absorption line than the Voigt profile and tends to yield a better fit. 

Using the Levenberg–Marquardt algorithm from the \texttt{lmfit} python package, we minimize $\chi^2$ between the observed set of coadded absorption lines over multiple exposures and the synthetic absorption lines with $\log g$, $T_{\text{eff}}$, and the apparent radial velocity ($v_r$) as free parameters. This creates a best-fit reference spectrum parameterized by $\log g$ and $T_{\text{eff}}$.

Then $v_r$ is measured by adding different radial velocities to the model and calculating $\chi^2$ between the observed and model spectrum at a grid ranging from $-1,500$~km s$^{-1}$ to $1,500$~km s$^{-1}$ in steps of $0.5$~km s$^{-1}$. In the immediate vicinity of the minimum $\chi^2$ value, $\chi^2$ as a function of radial velocity is well approximated by a parabola. This is true as long as the distribution in uncertainty of the measured spectrum is Gaussian. We measure the true radial velocity by fitting a parabola to the curve in a window of $100$~km s$^{-1}$ around the lowest $\chi^2$ radial velocity and finding that parabola's minimum point. Standard uncertainties are then determined by taking all radial velocities $v_r$ satisfying
\begin{equation}
    \left|\chi^2(v_r) - \text{min}(\chi^2(v_r))\right| < 1
\end{equation}
to be within $1\sigma$, as recommended by \cite{Chi2Paper}. 

\begin{figure}
    \centering
    \includegraphics[width=\columnwidth]{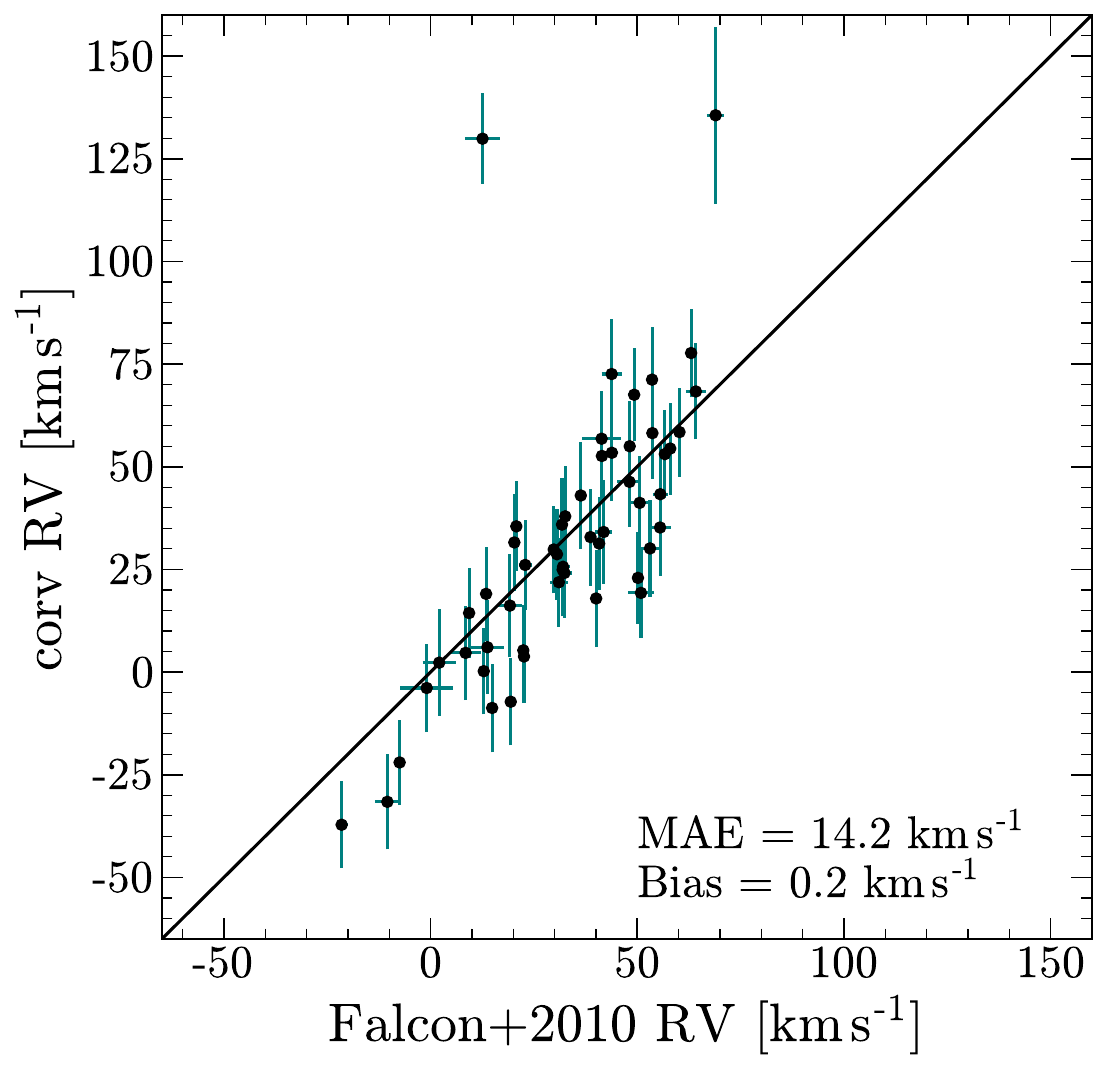}
    \caption{Radial velocities from \cite{Falcon_2010} plotted against those calculated in \texttt{corv}. We find 46 out of 53 white dwarfs consistent with \cite{Falcon_2010} to $2\sigma$, mean absolute error of $14.2$~km s$^{-1}$, and bias of $0.2$~km s$^{-1}$.}
    \label{fig:koester-falcon}
\end{figure}

We verify that \texttt{corv} yields reliable radial velocities by cross-matching the dataset of 3000 white dwarfs in SDSS-IV from \cite{Chandra_2020} with the high-resolution radial velocity measurements carried out by \cite{Falcon_2010} using data from the European Southern Observatory's SN Ia progenitor survey \citep{Napiwotzki}. Fifty-four stars are identified as being in both catalogs. We take the results of \cite{Falcon_2010} to be the `ground truth' for the overlapping objects: their high-resolution data reveal narrow cores in the white dwarf absorption lines which allow for a straightforward determination of the apparent radial velocities. For each of the stars in common, we compute radial velocities from SDSS spectra using \texttt{corv} and compare them to the radial velocities calculated by \cite{Falcon_2010}. Of the 53 white dwarfs, 46 are consistent with the theoretical model within $2\sigma$. This indicates that \texttt{corv} performs as expected. We find a mean absolute error of $14.2$~km s$^{-1}$, and a bias of $0.2$~km s$^{-1}$. This is within the radial velocity accuracy expected given the absolute wavelength calibration of SDSS-IV spectra. The radial velocities derived by \texttt{corv} are compared to those from \cite{Falcon_2010} in Figure \ref{fig:koester-falcon}.

\subsection{Main Sequence Radial Velocities}

The radial velocity of main sequence companions with SDSS spectra and without \textit{Gaia} DR3 radial velocities are obtained by cross-correlation of their spectra against the templates from the MaStar database \citep{Abdurrouf, yan_mastar}. MaStar is a library of stellar spectra observed by the MaNGA (Mapping Nearby Galaxies at Apache Point) survey \citep{Yan_2016, Bundy_2014}. These spectra are measured with a resolution of $R\sim 1,800$ in the wavelength range $3,622 \angstrom < \lambda < 10,354 \angstrom$. It contains spectra and measured stellar parameters of 59,266 main sequence stars from SDSS-IV. From these 59,266 spectra, we select 10,000 at random to serve as template spectra. 

For each main sequence star with an SDSS spectrum and no radial velocity in \textit{Gaia}, we compute $\chi^2$ against the chosen $10,000$ MaStar spectra in the region $5,000 \angstrom < \lambda < 9,000 \angstrom$. Spectra taken with the Apache Point Observatory are observed over a smaller range of wavelengths, so for these stars we use the range $6,000 \angstrom < \lambda < 7,000 \angstrom$. The MaStar spectrum with the smallest $\chi^2$ value is then taken as that main sequence star's reference spectrum. 

\begin{figure}[t!]
    \centering
    \includegraphics[width=\columnwidth]{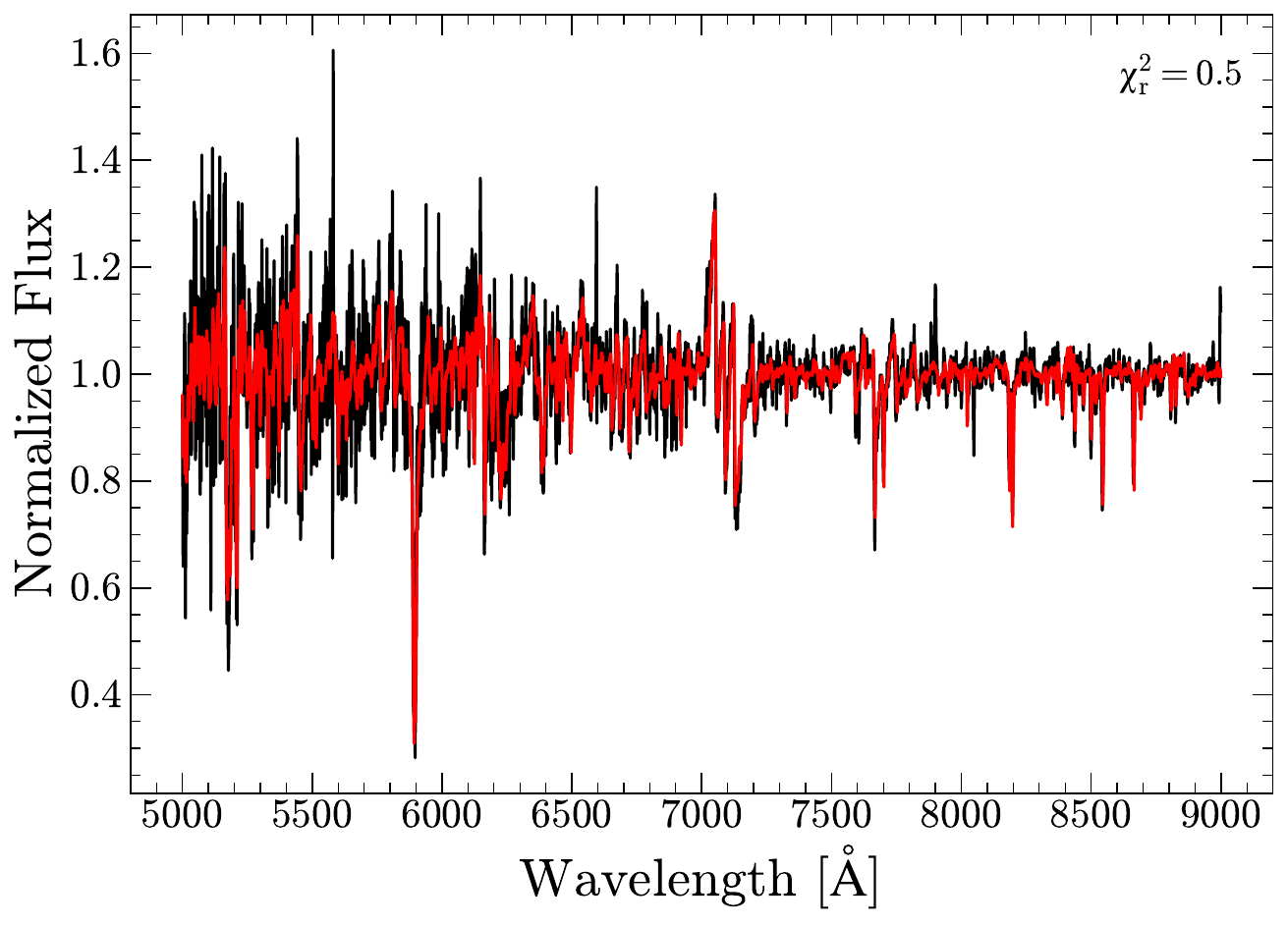}
    \caption{Main sequence radial velocity fit using templates from MaStar. Radial velocity is calculated from a template spectrum via $\chi^2$ minimization. \textit{Red:} MaStar template spectrum. \textit{Black:} Observed main sequence spectrum from SDSS-IV.}
    \label{fig:ms_rv}
\end{figure}

Then we compute radial velocities using $\chi^2$ minimization on a fine grid of radial velocities between $-1,500$~km s$^{-1}$ and $1,500$~km s$^{-1}$ in steps of $0.5$~km s$^{-1}$. This is the same method as that used by \texttt{corv} to calculate radial velocities from the best-fitting model spectrum from \cite{Koester2010}. Because main sequence spectra contain more absorption lines than white dwarfs and these lines are narrower and deeper, radial velocities can be measured to much higher accuracy. For this reason, we fit a parabola to the curve of $\chi^2$ as a function of radial velocity in a window of $5$~km s$^{-1}$ about the minimum $\chi^2$ radial velocity. Figure \ref{fig:ms_rv} presents an example of a MaStar template fit to an observed spectrum.

We validate these radial velocity fits by calculating radial velocities for 5,141 main sequence stars with spectra in SDSS and radial velocities in Gaia. This results in a mean average error of $14$~km s$^{-1}$ and bias of $-0.3$~km s$^{-1}$. 

Gravitational redshift from the surface of white dwarfs is then calculated as difference between the white dwarf's observed radial velocity and the main sequence star's radial velocity. Uncertainty is propagated as the sum in quadrature of the two radial velocities' uncertainties. 

We measure 305 gravitational redshifts by comparing the apparent velocities of the main sequence stars with the apparent velocities of the white dwarfs. The median separation of these wide binary pairs is $4,700$~AU, meaning that the orbital velocity is on the order of $0.1$~km s$^{-1}$, and can be neglected by comparison to other sources of error, the most significant of which is the error in radial velocity measurement of white dwarfs on the order of $15$ km s$^{-1}$. 


\begin{figure}[t!]
    \centering
    \includegraphics[width=\columnwidth]{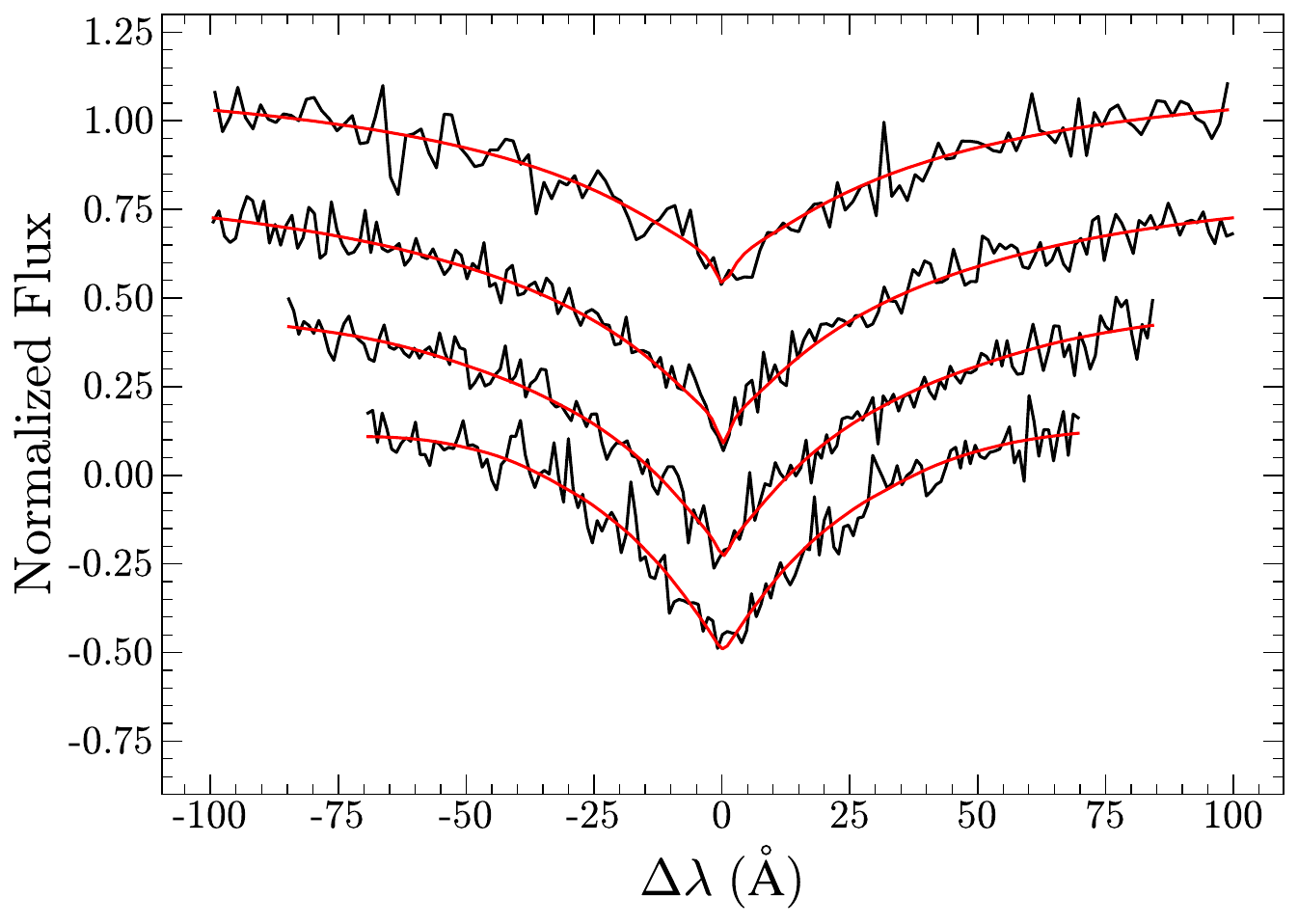}
    \caption{White dwarf radial velocity fit from \texttt{corv}. $T_{\text{eff}}$ and $\log g$ are fit using sample spectra from \cite{Koester2010}. Radial velocities are then fit to $H\alpha$, $H\beta$, $H\gamma$, and $H\delta$ lines simultaneously. Lines are continuum normalized and offset for clarity.}
    \label{fig:wd_rv}
\end{figure}

\subsection{Photometric Radii}

In order to determine the mass-radius relation of white dwarfs, a second constraint is needed in addition to the gravitational redshift. This can be obtained by directly measuring either mass or radius of the white dwarfs in the sample. In binaries with much smaller separations than those in our sample, the masses can sometimes be obtained from the dynamics of the orbit \citep{Bond_2017}, but these are not available in our case. Instead, we use an established procedure for calculating the radii of white dwarfs using their spectral energy distribution (SED) and the distance to the star. 

Surface flux for white dwarfs in a given band can be synthesized from $T_{\text{eff}}$ and $\log g$ using the latest synthetic DA spectra from \cite{synthspec}. We convolve these synthetic spectra into photometric flux in the observed band using the Python package \texttt{pyphot}\footnote{https://github.com/mfouesneau/pyphot}. The observed flux is then given by
\begin{equation}
    f_\nu = 4\pi \left(R/D\right)^2 H_\nu(T_{\text{eff}}, \log g),
\end{equation}
where $R$ is the radius of the star, $D$ is the distance to the star, and $H_\nu$ is the model surface flux for the chosen photometric band.

A degeneracy exists between radius and distance: a white dwarf that is twice as far away looks the same as another white dwarf with half the radius. Ignoring the orbital separation in comparison with the distance to the binaries, we use the distance to the main-sequence companions from \cite{Bailer_Jones_2021} since the companions are significantly brighter than the white dwarfs and therefore their parallaxes are measured with a higher signal-to-noise ratio. Since the distance is known, we can determine radius for the white dwarfs in the sample by fitting radius, $T_{\text{eff}}$, and $\log g$ to the photometry. For a uniform analysis, we use \textit{Gaia} photometry which is available for the entire sample instead of the SDSS photometry which is lacking for some targets. 

We correct for extinction due to interstellar dust using the extinction curve from \cite{FITZ}. Using the Python package \texttt{dustmaps} \citep{dustmaps}, we find $E(B-V)$ from the 3D dust map from \cite{bayestar}. We convert this into \textit{Gaia} magnitudes from the extinction curve with $R_V = 3.1$.





We fit the observed SEDs to the theoretical models by minimizing $\chi^2$ across \textit{Gaia} $G_\texttt{BP}$ and $G_\texttt{RP}$ photometric bands using the Levenberg-Marquardt algorithm as implemented in \texttt{lmfit}. $T_{\text{eff}}$ and radius freely vary in the fit, but we fix $\log g = 8$. The effect of $\log g$ on the photometry is negligible compared to that of $T_{\text{eff}}$ and radius and fixing it improves the convergence of the models to the observed flux. To account for the uncertainty added by fixing $\log g$, we also fit the SED using $\log g = 7$ and $\log g = 9$. Then we add the absolute difference of the radius values returned by those upper and lower radius bounds in quadrature with the standard deviation from the covariance matrix returned by the Levenberg-Marquart algorithm. This adds very small uncertainty to the result, with a median additional uncertainty of $0.3$\%. 

By fitting model atmospheres to the observed SED, we calculate the radii of 305 white dwarfs in the sample. Combined with the gravitational redshift measurements, these radii fully constrain the white dwarf mass-radius relation. 


\subsection{Sample Cleaning}\label{sec:cleaning}

\begin{figure}
    \centering
    \includegraphics[width=\columnwidth]{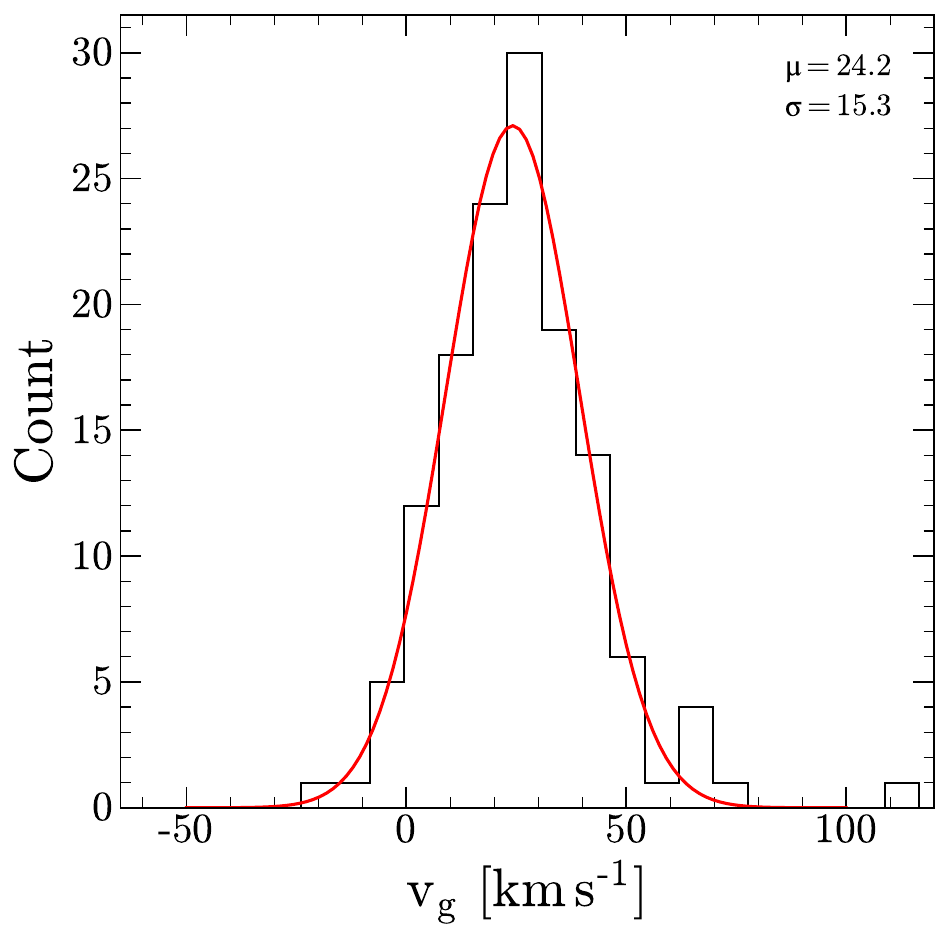}
    \caption{Sample distribution of white dwarf gravitational redshifts. The average gravitational redshift is $24.2$ km s$^{-1}$ with standard deviation $15.3$ km s$^{-1}$.}
    \label{fig:dist}
\end{figure}

Our final set of measurements of the gravitational redshifts and the radii still contains problematic data. Some spectra have a very low signal-to-noise ratio, causing the radial velocity fits to fail, or photometry or spectroscopy that has been contaminated by flux leakage due to nearby bright sources. In order to clean the sample, we first remove all data points whose white dwarf spectrum has signal-to-noise ratio less than $15$, using the medial signal to noise ratio in the wavelength range $5,000\,\angstrom < \lambda < 6,000\,\angstrom$. 

We next remove all targets whose white dwarf radial velocity fit has $\chi_r^2 > 5$ or whose main sequence radial velocity fit has $\chi_r^2 > 1.5$, if those main sequence radial velocities do not come from Gaia. These thresholds are determined by visually inspecting the fits. The sample is also restricted to only those white dwarfs whose measured radii are less than 0.0155 R$_{\odot}$. This is because any white dwarfs with larger radii are predicted to be in close binaries, whose apparent radial velocities can be contaminated by the orbital motion within the spatially unresolved binary \citep{Brown_2020}. One wide binary system is removed because a suitable template for the main sequence companion could not be identified, and the main sequence companion itself might be a double-lined binary. Finally, we remove all targets whose photometric radius fits have $\chi_r^2 > 5$, ensuring that all the targets in the final sample have accurate measurements. 


After applying these cuts, we are left with a final sample size of 137 stars. Of these, 87 white dwarf spectra come from SDSS-IV, 44 white dwarf spectra are from SDSS-V, and 6 radial velocities come from the \cite{Falcon_2010} catalog. One hundred and twelve main sequence radial velocities come from Gaia, one from SDSS-IV, 21 from SDSS-V, and 3 from the Apache Point Observatory. Figure \ref{fig:dist} shows the distribution of calculated white dwarf gravitational redshifts for the final sample. Of these, 4 are in wide binaries with separation less than 4 arcseconds. At these distances, flux leakage will start to affect measured photometric radii.

\section{Results} \label{sec:results}

\begin{table}
    \centering
    \begin{tabularx}{\columnwidth}{bss}
        \hline
        Radial Velocity From & WD & MS  \\
        \hline
        SDSS I-IV & 87 & 1 \\
        SDSS V & 44 & 21 \\
        \textit{Gaia} & - & 112 \\
        Apache Point & - & 3 \\
        \cite{Falcon_2010} & 6 & - \\
        \hline
        \textbf{Total} & \textbf{137} & \textbf{137}
    \end{tabularx}
    \caption{Sources of radial velocities for the final sample after quality assurance cuts are performed.}
    \label{tab:final_sample}
\end{table}

\begin{figure*}[t!]
    \centering
    \includegraphics[width=\textwidth]{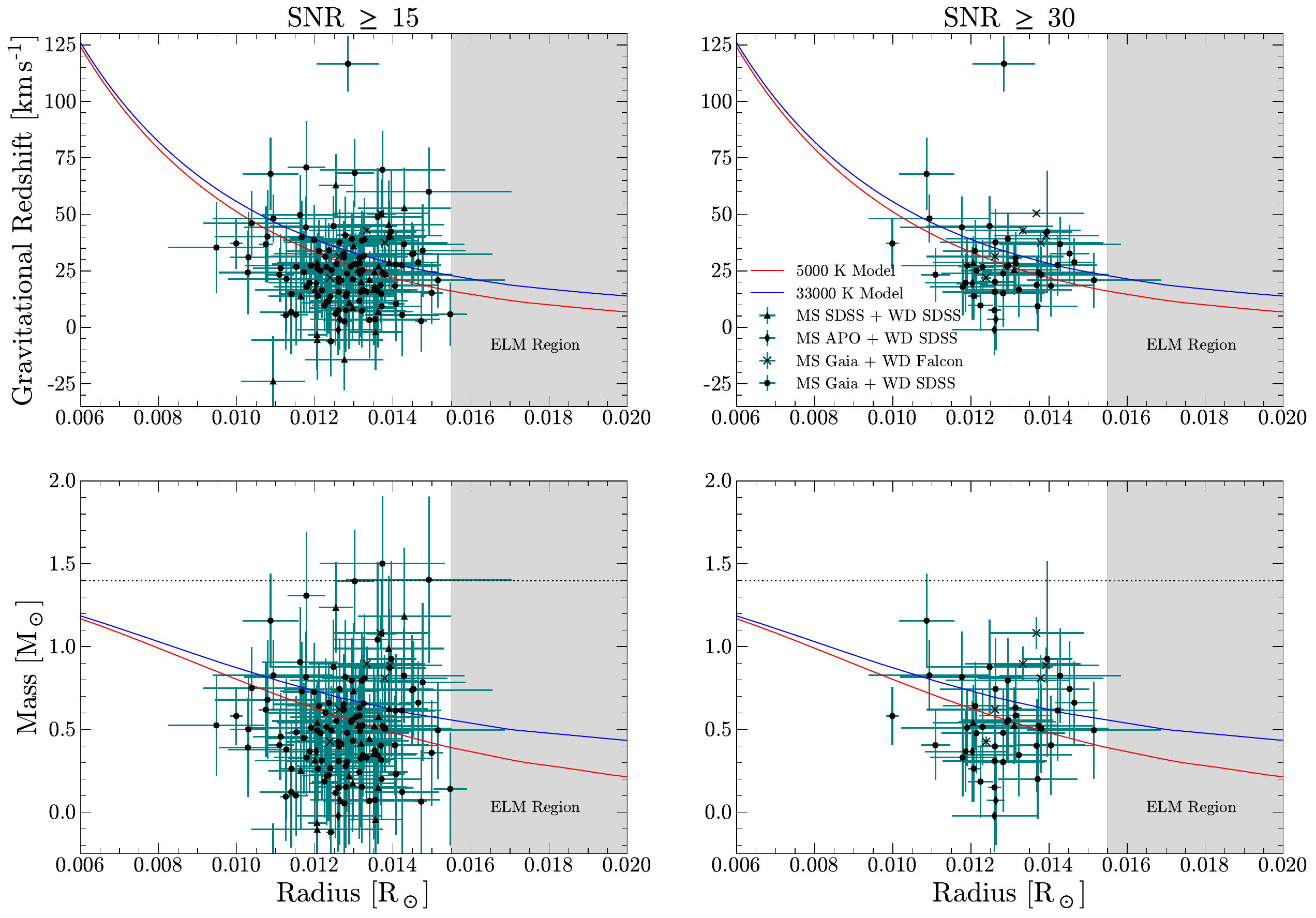}
    \caption{Gravitational redshift and mass as a function of radius. \textit{Left:} Gravitational redshift as a function of radius for all white dwarfs whose spectra have $\text{SNR} \geq 15$. The theoretical relation varies as a function of $T_{\text{eff}}$. The relation for the hottest stars in the sample ($33,000$ K) is represented in blue, and that for the coolest stars in the sample ($5,000$ K) is represented in red. \textit{Right:} The same, but restricted to white dwarfs with SDSS spectrum $\text{SNR} \geq 30$. Scatter in the original sample decreases when binaries whose white dwarf spectrum has $\text{SNR} < 30$ are removed. This restricts the sample to only those objects whose radial velocities are measured to the highest precision. The Chandrasekhar limit is marked in black on the bottom plots.}
    \label{fig:curve}
\end{figure*}

\begin{figure}
    \centering
    \includegraphics[width=\columnwidth]{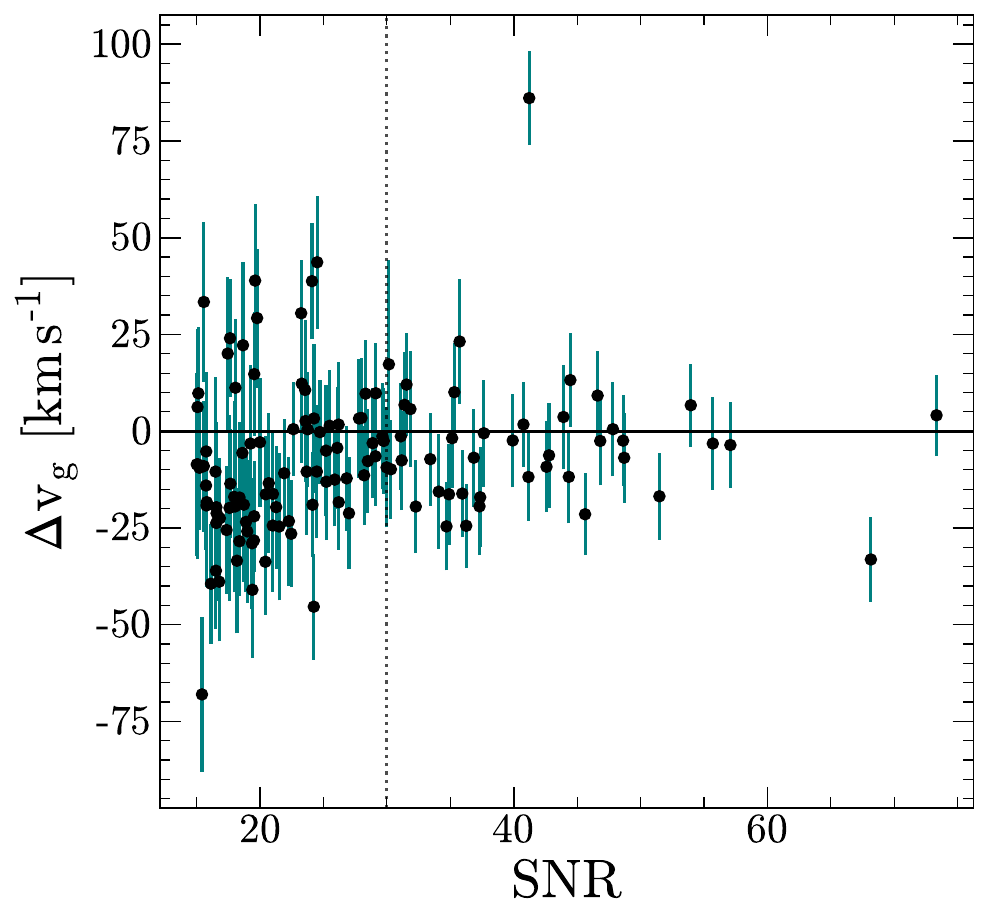}
    \caption{Difference between measured gravitational redshift and expected value for a white dwarf that's temperature is 14,000 K, corresponding to the mean sample temperature, plotted against spectrum signal-to-noise ratio (SNR). Low signal-to-noise ratio stars have more scatter and less consistency with the expected value than high signal-to-noise targets. A signal-to-noise ratio of 30 is marked as a horizontal line.}
    \label{fig:snr-vg}
\end{figure}

\begin{figure*}
    \centering
    \includegraphics[width=\textwidth]{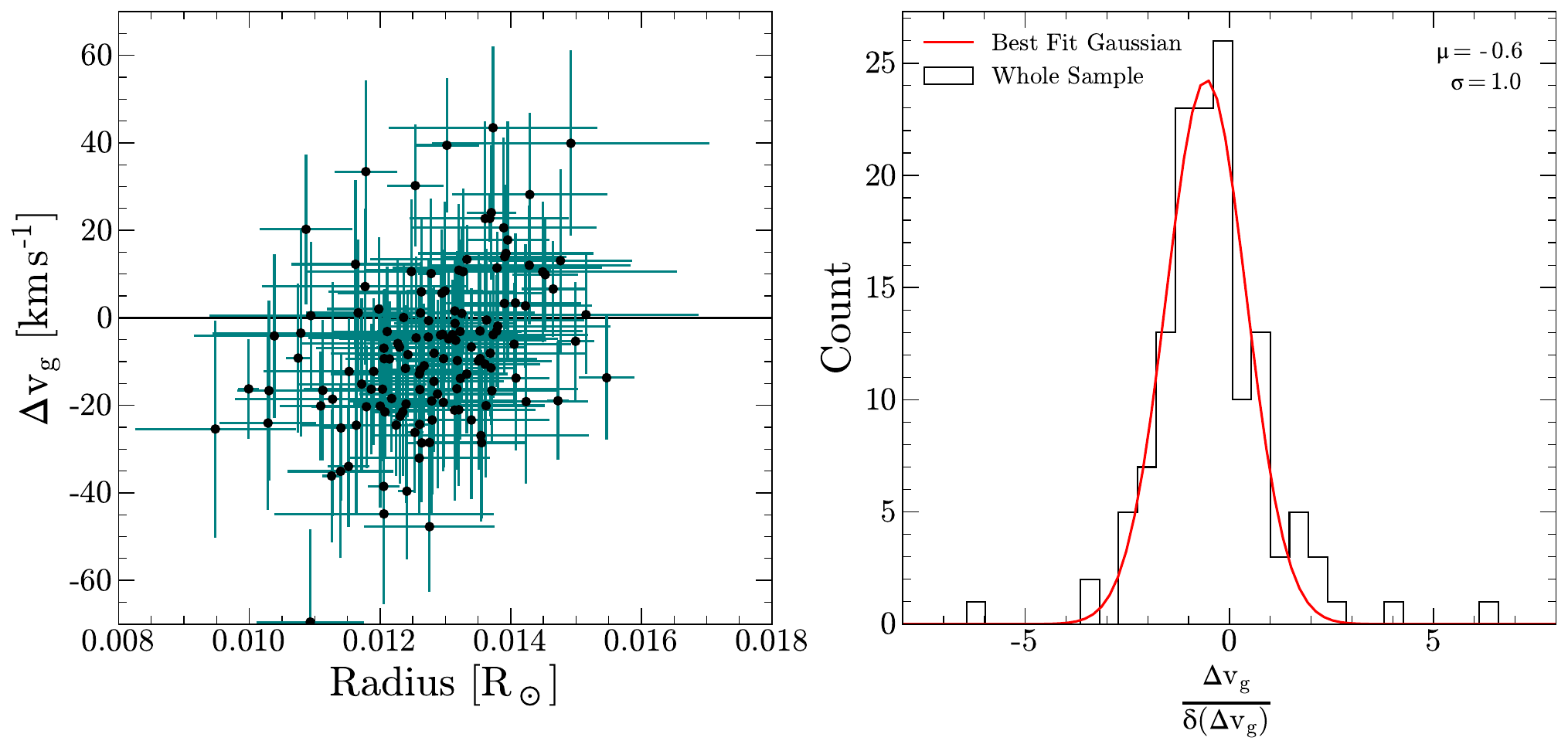}
    \caption{\textit{Left:} $\Delta v_g$ as a function of radius. The points cluster around the $\Delta v_g = 0$ line (black), indicating consistency between observation and the theoretical relations. $\Delta v_g$ is calculated using the theoretical model corresponding to the average temperature of the entire sample ($14,000$~K).\textit{Right:} Distribution of $\Delta v_g$ over error for each point. The fitted normal distribution (red) gives a sample average $\Delta v_g$ over error of $-0.5\pm 0.9$, indicating consistency with the theoretical model.}
    \label{fig:quantify}
\end{figure*}

\subsection{The Mass-Radius Relation}

Our final sample consists of 137 white dwarfs whose gravitational redshifts and radii are measured. This sample is described in Table \ref{tab:final_sample}. This sample is dominated by medium mass white dwarfs, reflecting the general distribution of masses of all white dwarfs. The bulk of the sample falls along the path of the theoretical relation. The spread of the measurements with respect to the theoretical relation is strongly dependent on the signal-to-noise ratio of the spectrum used to calculate the white dwarf companion's radial velocity. That is, restricting the sample to only those objects with high signal-to-noise ratio results in a closer correlation to the theoretical curve. Figure \ref{fig:curve} presents our sample and the theoretical relation for objects with $\text{SNR} \geq 15$ as well as $\text{SNR} \geq 30$. Figure \ref{fig:snr-vg} presents the deviation between measured gravitational redshift and the expected value for a white dwarf that's temperature is $14,000$ K, the temperature corresponding to the sample average, as a function of the spectrum's signal-to-noise ratio.

One notable outlier is present with $v_g = 116$ km s$^{-1}$ (\texttt{Gaia DR3 4005438916307756928}). This value of gravitational redshift is consistent with that measured by \cite{wegner}. They note the presence of hydrogen emission lines in the star's spectrum and that the star's radial velocity is possibly variable.

We quantitatively assess our results by computing $\Delta v_g$, the difference between each star's observed gravitational redshift and the theoretical gravitational redshift corresponding to $14,000$ K, which is near to the mean temperature of the entire sample. If our data are in good agreement with theoretical models, we should see that our data are consistent with $\Delta v_g = 0$. The steep dependency of gravitational redshift on radius can induce a bias on $\Delta v_g$. This occurs because the range of theoretical values of $v_g$ that the star could take is not symmetrically distributed due to the shape of the model curve. To mitigate this potential bias, we compute $\Delta v_g$ for each point by sampling $25$ points from the Gaussian distribution corresponding to that point's radius uncertainty and averaging $\Delta v_g$ over those values.

We calculate uncertainty in $\Delta v_g$ via standard error propagation. We also calculate a distribution of $\Delta v_g$ over error. The model is consistent with observation if this number is consistent with $0$. We find an average $\Delta v_g$ over error of $-0.5\pm 0.9$, indicating good agreement between observations and theory. Figure \ref{fig:quantify} presents $\Delta v_g$ as a function of radius as well as the distribution of $\Delta v_g$ over error for the whole sample. 


The median uncertainty in gravitational redshift is $13$~km s$^{-1}$, dominated by the uncertainty in the apparent radial velocity of the white dwarfs. The absolute wavelength calibration of SDSS spectra is  approximately $7$~km s$^{-1}$. The uncertainties of the radial velocities of the main sequence stars are generally better than those of the white dwarfs. The best main sequence radial velocities are those from \textit{Gaia}, which are selected to have uncertainties less than $7$~km s$^{-1}$ and make up the bulk of the sample, and the worst radial velocities are those from SDSS, whose uncertainties are similar to those of the white dwarfs. Our photometric radius measurements have a median uncertainty of $\approx 0.0009$~R$_\odot$. The dominant source of this uncertainty is likely \textit{Gaia} parallax and surface temperatures. 



\subsection{Temperature Variations}



\begin{figure}
    \centering
    \includegraphics[width=\columnwidth]{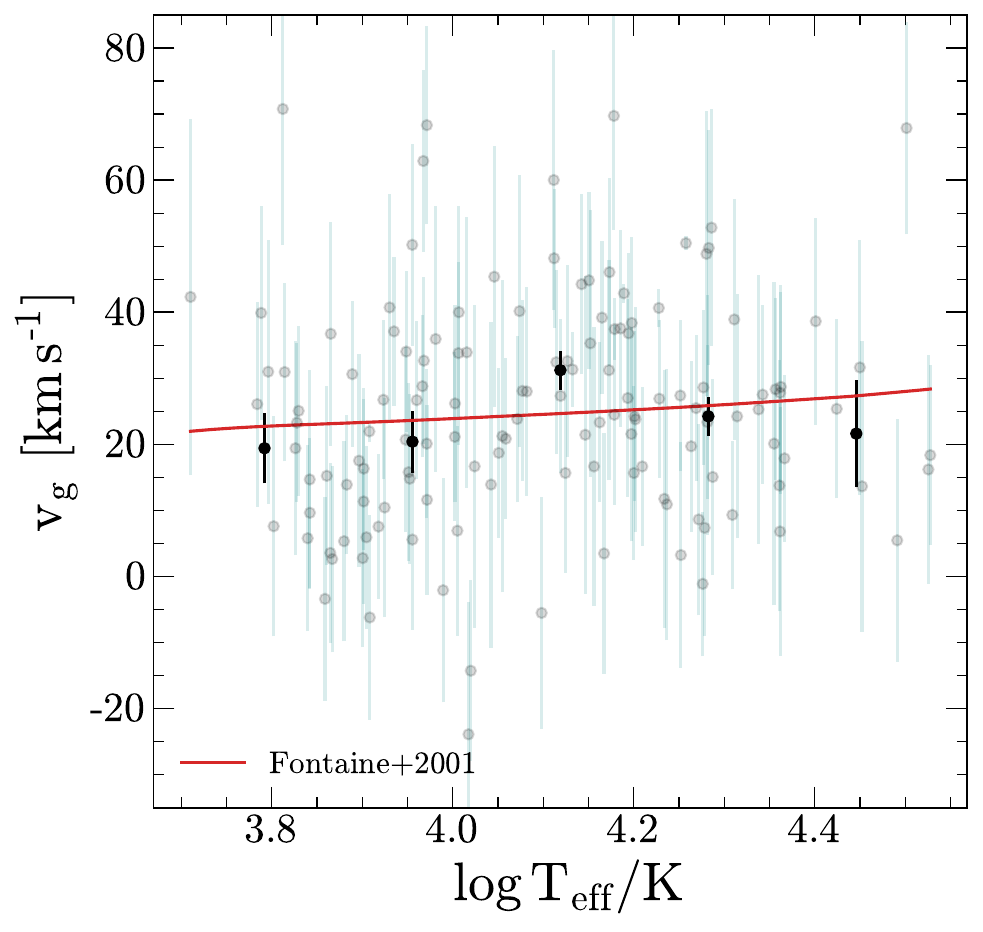}
    \caption{$v_g$ as a function of $\log T_{\text{eff}}$ for the whole sample. The sample is split into bins on $T_{\text{eff}}$ and the median $v_g$ is plotted for each bin (\textit{black}). It is clear that $v_g$ is consistent with the theoretical model. \textit{Red:} The theoretical relation predicted by \cite{Fontaine_2001}.}
    \label{fig:teff-deltavg}
\end{figure}




To understand the effect of $T_{\text{eff}}$ on a white dwarf's gravitational redshift, we plot the measured gravitational redshift of each white dwarf in the sample against its effective temperature as measured during the photometric radius fit. Figure \ref{fig:teff-deltavg} presents $v_g$ as a function of the logarithm of photometric $T_{\text{eff}}$ for each star in the final sample. The sample is split into $T_{\text{eff}}$ bins, and the median $v_g$ is computed for each bin. Although the error bars are larger, our observations are consistent with the model from \cite{Fontaine_2001}. It is clear that larger samples will be required to reduce the statistical uncertainty and measure the subtle temperature-dependence of the WD EoS.



\section{Summary and Conclusions} \label{sec:discussion}

We present a measurement of the theoretical white dwarf mass-radius relation based on the gravitational redshifts of individual white dwarfs in wide binaries with main sequence stars. We use a catalog of wide binary stars from \cite{El_Badry_2021}, astrometry and photometry from \textit{Gaia}, and spectroscopy from the Sloan Digital Sky Survey and the Apache Point Observatory to measure the individual gravitational redshifts of 137 white dwarfs, the largest set of such measurements to date. Combining gravitational redshifts with photometric temperatures and radii, we measure the mass-radius relation of white dwarfs. Our results are consistent with the relationship predicted by \cite{Fontaine_2001} (with average an $\Delta v_g$ over error of $-0.5 \pm 0.9$) and represent the largest catalog of individual gravitational redshift measurements independent of the mass-radius relation.

Some scatter in our data can be explained by unresolved close binaries in either companion. For such systems, the effect of orbital motion on radial velocities is no longer negligible. We control for this by removing low-mass white dwarfs whose evolutionary pathways make it probable that they are in close binaries, but this method cannot account for the possible binarity of the main sequence star. Thus, some scatter due to unresolved binaries is expected. Nonetheless, the scatter in our sample is consistent with errors, indicating that the impact of close binaries is small.

Although we constrain the relation for mid-mass white dwarfs well, our sample lacks observations of high-mass white dwarfs. It is only in the high mass regime that mass-radius relations due to core composition (e.g. C/O vs O/Ne) become apparent. In that regime, the expected gravitational redshifts differ by $~5$ km\,s$^{-1}$. High-resolution constraints on the gravitational redshift of high-mass white dwarfs are of great scientific interest and remain an outstanding area for future investigation.

Due to the limited resolution of the available spectra, our data cannot accurately measure variations in the mass-radius relation due to core composition. The limited resolution of SDSS spectra place an upper bound on the precision of gravitational redshift measurements at approximately $10$~km \,s$^{-1}$. In order to constrain core compositions on the gravitational redshift curve, it will be necessary to measure redshifts from high-resolution spectra. Our results are consistent with the theoretical effect of temperature on gravitational redshift, but are not conclusive due to the sizable statistical uncertainties in our measurement. High resolution observations will be necessary to draw more detailed conclusions about the exact dependence of the mass-radius relation on surface temperature.


In conclusion, we have presented the largest test of the mass-radius relation of white dwarfs using direct measurements of individual stars' gravitational redshifts. 
Our results demonstrate the utility of all-sky surveys for white dwarf science. With more spectra from future spectroscopic data releases, it will be possible to increase the sample size of this measurement by at least an order of magnitude, and perhaps characterize variations due to temperature and core composition.

\section*{Acknowledgements}

S.A. was supported by the JHU Provost's Undergraduate Research Award. N.L.Z. acknowledges support by the seed grant from the JHU Institute for Data Intensive Engineering and Science. V.C. acknowledges a Peirce Fellowship from Harvard University. 

Funding for the Sloan Digital Sky Survey V has been provided by the Alfred P. Sloan Foundation, the Heising-Simons Foundation, the National Science Foundation, and the Participating Institutions. SDSS acknowledges support and resources from the Center for High-Performance Computing at the University of Utah. The SDSS web site is \url{www.sdss.org}.

SDSS is managed by the Astrophysical Research Consortium for the Participating Institutions of the SDSS Collaboration, including the Carnegie Institution for Science, Chilean National Time Allocation Committee (CNTAC) ratified researchers, the Gotham Participation Group, Harvard University, Heidelberg University, The Johns Hopkins University, L’Ecole polytechnique f\'{e}d\'{e}rale de Lausanne (EPFL), Leibniz-Institut f\"{u}r Astrophysik Potsdam (AIP), Max-Planck-Institut f\"{u}r Astronomie (MPIA Heidelberg), Max-Planck-Institut f\"{u}r Extraterrestrische Physik (MPE), Nanjing University, National Astronomical Observatories of China (NAOC), New Mexico State University, The Ohio State University, Pennsylvania State University, Smithsonian Astrophysical Observatory, Space Telescope Science Institute (STScI), the Stellar Astrophysics Participation Group, Universidad Nacional Aut\'{o}noma de M\'{e}xico, University of Arizona, University of Colorado Boulder, University of Illinois at Urbana-Champaign, University of Toronto, University of Utah, University of Virginia, Yale University, and Yunnan University.

This work has made use of data from the European Space Agency (ESA) mission \textit{Gaia} (\url{https://www.cosmos.esa.int/gaia}), processed by the \textit{Gaia} Data Processing and Analysis Consortium (DPAC, \url{https://www.cosmos.esa.int/web/gaia/dpac/consortium}). Funding for the DPAC has been provided by national institutions, in particular the institutions participating in the \textit{Gaia} Multilateral Agreement.


\section*{Data availability}

The wavelength-calibrated and flux-calibrated SDSS spectra of white dwarfs and main sequence stars used in this paper (whether they are from proprietary SDSS-V data or public SDSS I-IV data) are available as an electronic supplement to this paper, with the wavelengths given in vacuum as per the SDSS convention. The wavelength-calibrated APO spectra used in this paper are also available, with the wavelengths given in air. APO flux calibrations should not be used due to continued issues with the instrument. 






\bibliography{citations.bib}{}

\begin{thebibliography}{}
\expandafter\ifx\csname natexlab\endcsname\relax\def\natexlab#1{#1}\fi
\providecommand{\url}[1]{\href{#1}{#1}}
\providecommand{\dodoi}[1]{doi:~\href{http://doi.org/#1}{\nolinkurl{#1}}}
\providecommand{\doeprint}[1]{\href{http://ascl.net/#1}{\nolinkurl{http://ascl.net/#1}}}
\providecommand{\doarXiv}[1]{\href{https://arxiv.org/abs/#1}{\nolinkurl{https://arxiv.org/abs/#1}}}

\bibitem[{{Abdurro'uf} {et~al.}(2022){Abdurro'uf}, {Accetta}, {Aerts}, {Silva
  Aguirre}, {Ahumada}, {Ajgaonkar}, {Filiz Ak}, {Alam}, {Allende Prieto},
  {Almeida}, {Anders}, {Anderson}, {Andrews}, {Anguiano}, {Aquino-Ort{\'\i}z},
  {Arag{\'o}n-Salamanca}, {Argudo-Fern{\'a}ndez}, {Ata}, {Aubert},
  {Avila-Reese}, {Badenes}, {Barb{\'a}}, {Barger}, {Barrera-Ballesteros},
  {Beaton}, {Beers}, {Belfiore}, {Bender}, {Bernardi}, {Bershady}, {Beutler},
  {Bidin}, {Bird}, {Bizyaev}, {Blanc}, {Blanton}, {Boardman}, {Bolton},
  {Boquien}, {Borissova}, {Bovy}, {Brandt}, {Brown}, {Brownstein}, {Brusa},
  {Buchner}, {Bundy}, {Burchett}, {Bureau}, {Burgasser}, {Cabang}, {Campbell},
  {Cappellari}, {Carlberg}, {Wanderley}, {Carrera}, {Cash}, {Chen}, {Chen},
  {Cherinka}, {Chiappini}, {Choi}, {Chojnowski}, {Chung}, {Clerc}, {Cohen},
  {Comerford}, {Comparat}, {da Costa}, {Covey}, {Crane}, {Cruz-Gonzalez},
  {Culhane}, {Cunha}, {Dai}, {Damke}, {Darling}, {Davidson}, {Davies},
  {Dawson}, {De Lee}, {Diamond-Stanic}, {Cano-D{\'\i}az}, {S{\'a}nchez},
  {Donor}, {Duckworth}, {Dwelly}, {Eisenstein}, {Elsworth}, {Emsellem},
  {Eracleous}, {Escoffier}, {Fan}, {Farr}, {Feng}, {Fern{\'a}ndez-Trincado},
  {Feuillet}, {Filipp}, {Fillingham}, {Frinchaboy}, {Fromenteau}, {Galbany},
  {Garc{\'\i}a}, {Garc{\'\i}a-Hern{\'a}ndez}, {Ge}, {Geisler}, {Gelfand},
  {G{\'e}ron}, {Gibson}, {Goddy}, {Godoy-Rivera}, {Grabowski}, {Green},
  {Greener}, {Grier}, {Griffith}, {Guo}, {Guy}, {Hadjara}, {Harding},
  {Hasselquist}, {Hayes}, {Hearty}, {Hern{\'a}ndez}, {Hill}, {Hogg},
  {Holtzman}, {Horta}, {Hsieh}, {Hsu}, {Hsu}, {Huber}, {Huertas-Company},
  {Hutchinson}, {Hwang}, {Ibarra-Medel}, {Chitham}, {Ilha}, {Imig}, {Jaekle},
  {Jayasinghe}, {Ji}, {Johnson}, {Jones}, {J{\"o}nsson}, {Katkov}, {Khalatyan},
  {Kinemuchi}, {Kisku}, {Knapen}, {Kneib}, {Kollmeier}, {Kong}, {Kounkel},
  {Kreckel}, {Krishnarao}, {Lacerna}, {Lane}, {Langgin}, {Lavender}, {Law},
  {Lazarz}, {Leung}, {Leung}, {Lewis}, {Li}, {Li}, {Lian}, {Liang}, {Lin},
  {Lin}, {Lin}, {Lintott}, {Long}, {Longa-Pe{\~n}a}, {L{\'o}pez-Cob{\'a}},
  {Lu}, {Lundgren}, {Luo}, {Mackereth}, {de la Macorra}, {Mahadevan},
  {Majewski}, {Manchado}, {Mandeville}, {Maraston}, {Margalef-Bentabol},
  {Masseron}, {Masters}, {Mathur}, {McDermid}, {Mckay}, {Merloni},
  {Merrifield}, {Meszaros}, {Miglio}, {Di Mille}, {Minniti}, {Minsley},
  {Monachesi}, {Moon}, {Mosser}, {Mulchaey}, {Muna}, {Mu{\~n}oz}, {Myers},
  {Myers}, {Nadathur}, {Nair}, {Nandra}, {Neumann}, {Newman}, {Nidever},
  {Nikakhtar}, {Nitschelm}, {O'Connell}, {Garma-Oehmichen}, {Luan Souza de
  Oliveira}, {Olney}, {Oravetz}, {Ortigoza-Urdaneta}, {Osorio}, {Otter},
  {Pace}, {Padilla}, {Pan}, {Pan}, {Parikh}, {Parker}, {Peirani}, {Pe{\~n}a
  Ram{\'\i}rez}, {Penny}, {Percival}, {Perez-Fournon}, {Pinsonneault},
  {Poidevin}, {Poovelil}, {Price-Whelan}, {B{\'a}rbara de Andrade Queiroz},
  {Raddick}, {Ray}, {Rembold}, {Riddle}, {Riffel}, {Riffel}, {Rix}, {Robin},
  {Rodr{\'\i}guez-Puebla}, {Roman-Lopes}, {Rom{\'a}n-Z{\'u}{\~n}iga}, {Rose},
  {Ross}, {Rossi}, {Rubin}, {Salvato}, {S{\'a}nchez}, {S{\'a}nchez-Gallego},
  {Sanderson}, {Santana Rojas}, {Sarceno}, {Sarmiento}, {Sayres}, {Sazonova},
  {Schaefer}, {Schiavon}, {Schlegel}, {Schneider}, {Schultheis}, {Schwope},
  {Serenelli}, {Serna}, {Shao}, {Shapiro}, {Sharma}, {Shen}, {Shetrone}, {Shu},
  {Simon}, {Skrutskie}, {Smethurst}, {Smith}, {Sobeck}, {Spoo}, {Sprague},
  {Stark}, {Stassun}, {Steinmetz}, {Stello}, {Stone-Martinez},
  {Storchi-Bergmann}, {Stringfellow}, {Stutz}, {Su}, {Taghizadeh-Popp},
  {Talbot}, {Tayar}, {Telles}, {Teske}, {Thakar}, {Theissen}, {Tkachenko},
  {Thomas}, {Tojeiro}, {Hernandez Toledo}, {Troup}, {Trump}, {Trussler},
  {Turner}, {Tuttle}, {Unda-Sanzana}, {V{\'a}zquez-Mata}, {Valentini},
  {Valenzuela}, {Vargas-Gonz{\'a}lez}, {Vargas-Maga{\~n}a}, {Alfaro},
  {Villanova}, {Vincenzo}, {Wake}, {Warfield}, {Washington}, {Weaver},
  {Weijmans}, {Weinberg}, {Weiss}, {Westfall}, {Wild}, {Wilde}, {Wilson},
  {Wilson}, {Wilson}, {Wolf}, {Wood-Vasey}, {Yan}, {Zamora}, {Zasowski},
  {Zhang}, {Zhao}, {Zheng}, {Zheng}, \& {Zhu}}]{Abdurrouf}
{Abdurro'uf}, {Accetta}, K., {Aerts}, C., {et~al.} 2022, \apjs, 259, 35,
  \dodoi{10.3847/1538-4365/ac4414}

\bibitem[{Avni(1976)}]{Chi2Paper}
Avni, Y. 1976, Astrophysical Journal, 210, 642, \dodoi{10.1086/154870}

\bibitem[{Bailer-Jones {et~al.}(2021)Bailer-Jones, Rybizki, Fouesneau,
  Demleitner, \& Andrae}]{Bailer_Jones_2021}
Bailer-Jones, C. A.~L., Rybizki, J., Fouesneau, M., Demleitner, M., \& Andrae,
  R. 2021, The Astronomical Journal, 161, 147, \dodoi{10.3847/1538-3881/abd806}

\bibitem[{{Bergeron} {et~al.}(1995){Bergeron}, {Liebert}, \&
  {Fulbright}}]{bergeron_gr}
{Bergeron}, P., {Liebert}, J., \& {Fulbright}, M.~S. 1995, \apj, 444, 810,
  \dodoi{10.1086/175654}

\bibitem[{{Bond} {et~al.}(2017){Bond}, {Bergeron}, \& {B{\'e}dard}}]{Bond_2017}
{Bond}, H.~E., {Bergeron}, P., \& {B{\'e}dard}, A. 2017, \apj, 848, 16,
  \dodoi{10.3847/1538-4357/aa8a63}

\bibitem[{{Bowen} \& {Vaughan}(1973)}]{bowen_1973}
{Bowen}, I.~S., \& {Vaughan}, A.~H., J. 1973, \ao, 12, 1430,
  \dodoi{10.1364/AO.12.001430}

\bibitem[{Brown {et~al.}(2020)Brown, Kilic, Kosakowski, Andrews, Heinke,
  Agüeros, Camilo, Gianninas, Hermes, \& Kenyon}]{Brown_2020}
Brown, W.~R., Kilic, M., Kosakowski, A., {et~al.} 2020, The Astrophysical
  Journal, 889, 49, \dodoi{10.3847/1538-4357/ab63cd}

\bibitem[{Bundy {et~al.}(2014)Bundy, Bershady, Law, Yan, Drory, MacDonald,
  Wake, Cherinka, S{\'{a}}nchez-Gallego, Weijmans, Thomas, Tremonti, Masters,
  Coccato, Diamond-Stanic, Arag{\'{o}}n-Salamanca, Avila-Reese, Badenes,
  Falc{\'{o}}n-Barroso, Belfiore, Bizyaev, Blanc, Bland-Hawthorn, Blanton,
  Brownstein, Byler, Cappellari, Conroy, Dutton, Emsellem, Etherington,
  Frinchaboy, Fu, Gunn, Harding, Johnston, Kauffmann, Kinemuchi, Klaene,
  Knapen, Leauthaud, Li, Lin, Maiolino, Malanushenko, Malanushenko, Mao,
  Maraston, McDermid, Merrifield, Nichol, Oravetz, Pan, Parejko, Sanchez,
  Schlegel, Simmons, Steele, Steinmetz, Thanjavur, Thompson, Tinker, van~den
  Bosch, Westfall, Wilkinson, Wright, Xiao, \& Zhang}]{Bundy_2014}
Bundy, K., Bershady, M.~A., Law, D.~R., {et~al.} 2014, The Astrophysical
  Journal, 798, 7, \dodoi{10.1088/0004-637x/798/1/7}

\bibitem[{{Camisassa} {et~al.}(2019){Camisassa}, {Althaus}, {C{\'o}rsico}, {De
  Ger{\'o}nimo}, {Miller Bertolami}, {Novarino}, {Rohrmann}, {Wachlin}, \&
  {Garc{\'\i}a-Berro}}]{refId0}
{Camisassa}, M.~E., {Althaus}, L.~G., {C{\'o}rsico}, A.~H., {et~al.} 2019,
  \aap, 625, A87, \dodoi{10.1051/0004-6361/201833822}

\bibitem[{{Chandra} {et~al.}(2020){Chandra}, {Hwang}, {Zakamska}, \&
  {Budav{\'a}ri}}]{chandra_2020_wdtools}
{Chandra}, V., {Hwang}, H.-C., {Zakamska}, N.~L., \& {Budav{\'a}ri}, T. 2020,
  \mnras, 497, 2688, \dodoi{10.1093/mnras/staa2165}

\bibitem[{Chandra {et~al.}(2020)Chandra, Hwang, Zakamska, \&
  Cheng}]{Chandra_2020}
Chandra, V., Hwang, H.-C., Zakamska, N.~L., \& Cheng, S. 2020, The
  Astrophysical Journal, 899, 146, \dodoi{10.3847/1538-4357/aba8a2}

\bibitem[{Chandrasekhar \& Milne(1933)}]{10.1093/mnras/93.5.390}
Chandrasekhar, S., \& Milne, E.~A. 1933, Monthly Notices of the Royal
  Astronomical Society, 93, 390, \dodoi{10.1093/mnras/93.5.390}

\bibitem[{Cukanovaite {et~al.}(2023)Cukanovaite, Tremblay, Toonen, Temmink,
  Manser, O'Brien, \& McCleery}]{Cukanovaite_2023}
Cukanovaite, E., Tremblay, P.-E., Toonen, S., {et~al.} 2023, Monthly Notices of
  the Royal Astronomical Society, 522, 1643, \dodoi{10.1093/mnras/stad1020}

\bibitem[{Einstein(1916)}]{https://doi.org/10.1002/andp.19163540702}
Einstein, A. 1916, Annalen der Physik, 354, 769,
  \dodoi{https://doi.org/10.1002/andp.19163540702}

\bibitem[{El-Badry(2022)}]{elbadry2022gravitational}
El-Badry, K. 2022, The gravitational redshift of solar-type stars from Gaia DR3
  wide binaries.
\newblock \doarXiv{2206.11092}

\bibitem[{El-Badry {et~al.}(2021)El-Badry, Rix, \& Heintz}]{El_Badry_2021}
El-Badry, K., Rix, H.-W., \& Heintz, T.~M. 2021, Monthly Notices of the Royal
  Astronomical Society, 506, 2269, \dodoi{10.1093/mnras/stab323}

\bibitem[{Falcon {et~al.}(2010)Falcon, Winget, Montgomery, \&
  Williams}]{Falcon_2010}
Falcon, R.~E., Winget, D.~E., Montgomery, M.~H., \& Williams, K.~A. 2010, The
  Astrophysical Journal, 712, 585, \dodoi{10.1088/0004-637X/712/1/585}

\bibitem[{{Fitzpatrick}(1999)}]{FITZ}
{Fitzpatrick}, E.~L. 1999, \pasp, 111, 63, \dodoi{10.1086/316293}

\bibitem[{Fontaine {et~al.}(2001)Fontaine, Brassard, \&
  Bergeron}]{Fontaine_2001}
Fontaine, G., Brassard, P., \& Bergeron, P. 2001, Publications of the
  Astronomical Society of the Pacific, 113, 409, \dodoi{10.1086/319535}

\bibitem[{{Green}(2018)}]{dustmaps}
{Green}, G. 2018, The Journal of Open Source Software, 3, 695,
  \dodoi{10.21105/joss.00695}

\bibitem[{{Green} {et~al.}(2019){Green}, {Schlafly}, {Zucker}, {Speagle}, \&
  {Finkbeiner}}]{bayestar}
{Green}, G.~M., {Schlafly}, E., {Zucker}, C., {Speagle}, J.~S., \&
  {Finkbeiner}, D. 2019, \apj, 887, 93, \dodoi{10.3847/1538-4357/ab5362}

\bibitem[{{Greenstein} \& {Trimble}(1972)}]{greenstein_gr2}
{Greenstein}, J.~L., \& {Trimble}, V. 1972, \apjl, 175, L1,
  \dodoi{10.1086/180973}

\bibitem[{{Greenstein} \& {Trimble}(1967{\natexlab{a}})}]{greenstein_gr}
{Greenstein}, J.~L., \& {Trimble}, V.~L. 1967{\natexlab{a}}, \apj, 149, 283,
  \dodoi{10.1086/149254}

\bibitem[{{Greenstein} \& {Trimble}(1967{\natexlab{b}})}]{greenstein}
---. 1967{\natexlab{b}}, \apj, 149, 283, \dodoi{10.1086/149254}

\bibitem[{{Gunn} {et~al.}(2006){Gunn}, {Siegmund}, {Mannery}, {Owen}, {Hull},
  {Leger}, {Carey}, {Knapp}, {York}, {Boroski}, {Kent}, {Lupton}, {Rockosi},
  {Evans}, {Waddell}, {Anderson}, {Annis}, {Barentine}, {Bartoszek}, {Bastian},
  {Bracker}, {Brewington}, {Briegel}, {Brinkmann}, {Brown}, {Carr},
  {Czarapata}, {Drennan}, {Dombeck}, {Federwitz}, {Gillespie}, {Gonzales},
  {Hansen}, {Harvanek}, {Hayes}, {Jordan}, {Kinney}, {Klaene}, {Kleinman},
  {Kron}, {Kresinski}, {Lee}, {Limmongkol}, {Lindenmeyer}, {Long}, {Loomis},
  {McGehee}, {Mantsch}, {Neilsen}, {Neswold}, {Newman}, {Nitta}, {Peoples},
  {Pier}, {Prieto}, {Prosapio}, {Rivetta}, {Schneider}, {Snedden}, \&
  {Wang}}]{gunn_2006}
{Gunn}, J.~E., {Siegmund}, W.~A., {Mannery}, E.~J., {et~al.} 2006, \aj, 131,
  2332, \dodoi{10.1086/500975}

\bibitem[{{Hansen} {et~al.}(2002){Hansen}, {Brewer}, {Fahlman}, {Gibson},
  {Ibata}, {Limongi}, {Rich}, {Richer}, {Shara}, \& {Stetson}}]{hansen_2002}
{Hansen}, B. M.~S., {Brewer}, J., {Fahlman}, G.~G., {et~al.} 2002, \apjl, 574,
  L155, \dodoi{10.1086/342528}

\bibitem[{{Hillebrandt} \& {Niemeyer}(2000)}]{hillebrandt_2000}
{Hillebrandt}, W., \& {Niemeyer}, J.~C. 2000, \araa, 38, 191,
  \dodoi{10.1146/annurev.astro.38.1.191}

\bibitem[{{Hwang} {et~al.}(2020){Hwang}, {Hamer}, {Zakamska}, \&
  {Schlaufman}}]{Hwang2020}
{Hwang}, H.-C., {Hamer}, J.~H., {Zakamska}, N.~L., \& {Schlaufman}, K.~C. 2020,
  \mnras, 497, 2250, \dodoi{10.1093/mnras/staa2124}

\bibitem[{{Hwang} {et~al.}(2021){Hwang}, {Ting}, {Schlaufman}, {Zakamska}, \&
  {Wyse}}]{Hwang2021}
{Hwang}, H.-C., {Ting}, Y.-S., {Schlaufman}, K.~C., {Zakamska}, N.~L., \&
  {Wyse}, R. F.~G. 2021, \mnras, 501, 4329, \dodoi{10.1093/mnras/staa3854}

\bibitem[{Joyce {et~al.}(2018)Joyce, Barstow, Holberg, Bond, Casewell, \&
  Burleigh}]{10.1093/mnras/sty2404}
Joyce, S. R.~G., Barstow, M.~A., Holberg, J.~B., {et~al.} 2018, Monthly Notices
  of the Royal Astronomical Society, 481, 2361, \dodoi{10.1093/mnras/sty2404}

\bibitem[{Katz {et~al.}(2023)Katz, Sartoretti, Guerrier, Panuzzo, Seabroke,
  Th{\'{e} }venin, Cropper, Benson, Blomme, Haigron, Marchal, Smith, Baker,
  Chemin, Damerdji, David, Dolding, Fr{\'{e}}mat, Gosset, Jan{\ss}en,
  Jasniewicz, Lobel, Plum, Samaras, Snaith, Soubiran, Vanel, Zwitter, Antoja,
  Arenou, Babusiaux, Brouillet, Caffau, Matteo, Fabre, Fabricius, Fragkoudi,
  Haywood, Huckle, Hottier, Lasne, Leclerc, Mastrobuono-Battisti, Royer,
  Teyssier, Zorec, Crifo, Piccolo, Turon, \& Viala}]{Katz_2023}
Katz, D., Sartoretti, P., Guerrier, A., {et~al.} 2023, \aap, 674, A5,
  \dodoi{10.1051/0004-6361/202244220}

\bibitem[{{Koester}(1987)}]{koester_gr}
{Koester}, D. 1987, \apj, 322, 852, \dodoi{10.1086/165779}

\bibitem[{Koester(2010)}]{Koester2010}
Koester, D. 2010, Memorie della Societa Astronomica Italiana, 81, 921

\bibitem[{{Kollmeier} {et~al.}(2019){Kollmeier}, {Anderson}, {Blanc},
  {Blanton}, {Covey}, {Crane}, {Drory}, {Frinchaboy}, {Froning}, {Johnson},
  {Kneib}, {Kreckel}, {Merloni}, {Pellegrini}, {Pogge}, {Ramirez}, {Rix},
  {Sayres}, {S{\'a}nchez-Gallego}, {Shen}, {Tkachenko}, {Trump}, {Tuttle},
  {Weijmans}, {Zasowski}, {Barbuy}, {Beaton}, {Bergemann}, {Bochanski},
  {Brandt}, {Casey}, {Cherinka}, {Eracleous}, {Fan}, {Garc{\'\i}a}, {Green},
  {Hekker}, {Lane}, {Longa-Pe{\~n}a}, {Mathur}, {Meza}, {Minchev}, {Myers},
  {Nidever}, {Nitschelm}, {O'Connell}, {Price-Whelan}, {Raddick}, {Rossi},
  {Sankrit}, {Simon}, {Stutz}, {Ting}, {Trakhtenbrot}, {Weaver}, {Willmer}, \&
  {Weinberg}}]{sdss-v}
{Kollmeier}, J., {Anderson}, S.~F., {Blanc}, G.~A., {et~al.} 2019, in Bulletin
  of the American Astronomical Society, Vol.~51, 274

\bibitem[{{Napiwotzki} {et~al.}(2001){Napiwotzki}, {Christlieb}, {Drechsel},
  {Hagen}, {Heber}, {Homeier}, {Karl}, {Koester}, {Leibundgut}, {Marsh},
  {Moehler}, {Nelemans}, {Pauli}, {Reimers}, {Renzini}, \&
  {Yungelson}}]{Napiwotzki}
{Napiwotzki}, R., {Christlieb}, N., {Drechsel}, H., {et~al.} 2001,
  Astronomische Nachrichten, 322, 411,
  \dodoi{10.1002/1521-3994(200112)322:5/6<411::AID-ASNA411>3.0.CO;2-I}

\bibitem[{{Oswalt} {et~al.}(1996){Oswalt}, {Smith}, {Wood}, \&
  {Hintzen}}]{oswalt_1996}
{Oswalt}, T.~D., {Smith}, J.~A., {Wood}, M.~A., \& {Hintzen}, P. 1996, \nat,
  382, 692, \dodoi{10.1038/382692a0}

\bibitem[{{Parsons} {et~al.}(2017){Parsons}, {G{\"a}nsicke}, {Marsh}, {Ashley},
  {Bours}, {Breedt}, {Burleigh}, {Copperwheat}, {Dhillon}, {Green}, {Hardy},
  {Hermes}, {Irawati}, {Kerry}, {Littlefair}, {McAllister}, {Rattanasoon},
  {Rebassa-Mansergas}, {Sahman}, \& {Schreiber}}]{parsons_2017}
{Parsons}, S.~G., {G{\"a}nsicke}, B.~T., {Marsh}, T.~R., {et~al.} 2017, \mnras,
  470, 4473, \dodoi{10.1093/mnras/stx1522}

\bibitem[{Pasquini {et~al.}(2019)Pasquini, Pala, Ludwig, Lẽao, de~Medeiros,
  \& Weiss}]{Pasquini_2019}
Pasquini, L., Pala, A.~F., Ludwig, H.-G., {et~al.} 2019, \aap, 627, L8,
  \dodoi{10.1051/0004-6361/201935835}

\bibitem[{Perlmutter {et~al.}(1997)Perlmutter, Gabi, Goldhaber, Goobar, Groom,
  Hook, Kim, Kim, Lee, Pain, Pennypacker, Small, Ellis, McMahon, Boyle,
  Bunclark, Carter, Irwin, Glazebrook, Newberg, Filippenko, Matheson, Dopita,
  \& and}]{Perlmutter_1997}
Perlmutter, S., Gabi, S., Goldhaber, G., {et~al.} 1997, The Astrophysical
  Journal, 483, 565, \dodoi{10.1086/304265}

\bibitem[{{Provencal} {et~al.}(1998){Provencal}, {Shipman}, {H{\o}g}, \&
  {Thejll}}]{Provencal_1998}
{Provencal}, J.~L., {Shipman}, H.~L., {H{\o}g}, E., \& {Thejll}, P. 1998, \apj,
  494, 759, \dodoi{10.1086/305238}

\bibitem[{{Reid}(1996)}]{reid_gr}
{Reid}, I.~N. 1996, \aj, 111, 2000, \dodoi{10.1086/117936}

\bibitem[{{Riess} {et~al.}(1998){Riess}, {Filippenko}, {Challis},
  {Clocchiatti}, {Diercks}, {Garnavich}, {Gilliland}, {Hogan}, {Jha},
  {Kirshner}, {Leibundgut}, {Phillips}, {Reiss}, {Schmidt}, {Schommer},
  {Smith}, {Spyromilio}, {Stubbs}, {Suntzeff}, \& {Tonry}}]{reiss_1998}
{Riess}, A.~G., {Filippenko}, A.~V., {Challis}, P., {et~al.} 1998, \aj, 116,
  1009, \dodoi{10.1086/300499}

\bibitem[{Smee {et~al.}(2013)Smee, Gunn, Uomoto, Roe, Schlegel, Rockosi, Carr,
  Leger, Dawson, Olmstead, Brinkmann, Owen, Barkhouser, Honscheid, Harding,
  Long, Lupton, Loomis, Anderson, Annis, Bernardi, Bhardwaj, Bizyaev, Bolton,
  Brewington, Briggs, Burles, Burns, Castander, Connolly, Davenport, Ebelke,
  Epps, Feldman, Friedman, Frieman, Heckman, Hull, Knapp, Lawrence, Loveday,
  Mannery, Malanushenko, Malanushenko, Merrelli, Muna, Newman, Nichol, Oravetz,
  Pan, Pope, Ricketts, Shelden, Sandford, Siegmund, Simmons, Smith, Snedden,
  Schneider, SubbaRao, Tremonti, Waddell, \& York}]{Smee_2013}
Smee, S.~A., Gunn, J.~E., Uomoto, A., {et~al.} 2013, The Astronomical Journal,
  146, 32, \dodoi{10.1088/0004-6256/146/2/32}

\bibitem[{{Tremblay} {et~al.}(2015){Tremblay}, {Gianninas}, {Kilic}, {Ludwig},
  {Steffen}, {Freytag}, \& {Hermes}}]{synthspec}
{Tremblay}, P.~E., {Gianninas}, A., {Kilic}, M., {et~al.} 2015, \apj, 809, 148,
  \dodoi{10.1088/0004-637X/809/2/148}

\bibitem[{Tremblay {et~al.}(2016)Tremblay, Gentile-Fusillo, Raddi, Jordan,
  Besson, Gänsicke, Parsons, Koester, Marsh, Bohlin, Kalirai, \&
  Deustua}]{tremblay_dr1}
Tremblay, P.-E., Gentile-Fusillo, N., Raddi, R., {et~al.} 2016, Monthly Notices
  of the Royal Astronomical Society, 465, 2849, \dodoi{10.1093/mnras/stw2854}

\bibitem[{{Wegner} \& {Reid}(1991)}]{wegner}
{Wegner}, G., \& {Reid}, I.~N. 1991, \apj, 375, 674, \dodoi{10.1086/170230}

\bibitem[{{Wilson} {et~al.}(2019){Wilson}, {Hearty}, {Skrutskie}, {Majewski},
  {Holtzman}, {Eisenstein}, {Gunn}, {Blank}, {Henderson}, {Smee}, {Nelson},
  {Nidever}, {Arns}, {Barkhouser}, {Barr}, {Beland}, {Bershady}, {Blanton},
  {Brunner}, {Burton}, {Carey}, {Carr}, {Colque}, {Crane}, {Damke}, {Davidson},
  {Dean}, {Di Mille}, {Don}, {Ebelke}, {Evans}, {Fitzgerald}, {Gillespie},
  {Hall}, {Harding}, {Harding}, {Hammond}, {Hancock}, {Harrison}, {Hope},
  {Horne}, {Karakla}, {Lam}, {Leger}, {MacDonald}, {Maseman}, {Matsunari},
  {Melton}, {Mitcheltree}, {O'Brien}, {O'Connell}, {Patten}, {Richardson},
  {Rieke}, {Rieke}, {Roman-Lopes}, {Schiavon}, {Sobeck}, {Stolberg}, {Stoll},
  {Tembe}, {Trujillo}, {Uomoto}, {Vernieri}, {Walker}, {Weinberg}, {Young},
  {Anthony-Brumfield}, {Bizyaev}, {Breslauer}, {De Lee}, {Downey}, {Halverson},
  {Huehnerhoff}, {Klaene}, {Leon}, {Long}, {Mahadevan}, {Malanushenko},
  {Nguyen}, {Owen}, {S{\'a}nchez-Gallego}, {Sayres}, {Shane}, {Shectman},
  {Shetrone}, {Skinner}, {Stauffer}, \& {Zhao}}]{wilson_2019}
{Wilson}, J.~C., {Hearty}, F.~R., {Skrutskie}, M.~F., {et~al.} 2019, \pasp,
  131, 055001, \dodoi{10.1088/1538-3873/ab0075}

\bibitem[{{Wood}(1992)}]{wood_1992}
{Wood}, M.~A. 1992, \apj, 386, 539, \dodoi{10.1086/171038}

\bibitem[{Yan {et~al.}(2016)Yan, Bundy, Law, Bershady, Andrews, Cherinka,
  Diamond-Stanic, Drory, MacDonald, S{\'{a}}nchez-Gallego, Thomas, Wake,
  Weijmans, Westfall, Zhang, Arag{\'{o}}n-Salamanca, Belfiore, Bizyaev, Blanc,
  Blanton, Brownstein, Cappellari, D'Souza, Emsellem, Fu, Gaulme, Graham,
  Goddard, Gunn, Harding, Jones, Kinemuchi, Li, Li, Maiolino, Mao, Maraston,
  Masters, Merrifield, Oravetz, Pan, Parejko, Sanchez, Schlegel, Simmons,
  Thanjavur, Tinker, Tremonti, van~den Bosch, \& Zheng}]{Yan_2016}
Yan, R., Bundy, K., Law, D.~R., {et~al.} 2016, The Astronomical Journal, 152,
  197, \dodoi{10.3847/0004-6256/152/6/197}

\bibitem[{{Yan} {et~al.}(2019){Yan}, {Chen}, {Lazarz}, {Bizyaev}, {Maraston},
  {Stringfellow}, {McCarthy}, {Meneses-Goytia}, {Law}, {Thomas}, {Falcon
  Barroso}, {S{\'a}nchez-Gallego}, {Schlafly}, {Zheng}, {Argudo-Fern{\'a}ndez},
  {Beaton}, {Beers}, {Bershady}, {Blanton}, {Brownstein}, {Bundy}, {Chambers},
  {Cherinka}, {De Lee}, {Drory}, {Galbany}, {Holtzman}, {Imig}, {Kaiser},
  {Kinemuchi}, {Liu}, {Luo}, {Magnier}, {Majewski}, {Nair}, {Oravetz},
  {Oravetz}, {Pan}, {Sobeck}, {Stassun}, {Talbot}, {Tremonti}, {Waters},
  {Weijmans}, {Wilhelm}, {Zasowski}, {Zhao}, \& {Zhao}}]{yan_mastar}
{Yan}, R., {Chen}, Y., {Lazarz}, D., {et~al.} 2019, \apj, 883, 175,
  \dodoi{10.3847/1538-4357/ab3ebc}

\bibitem[{{York} {et~al.}(2000){York}, {Adelman}, {Anderson}, {Anderson},
  {Annis}, {Bahcall}, {Bakken}, {Barkhouser}, {Bastian}, {Berman}, {Boroski},
  {Bracker}, {Briegel}, {Briggs}, {Brinkmann}, {Brunner}, {Burles}, {Carey},
  {Carr}, {Castander}, {Chen}, {Colestock}, {Connolly}, {Crocker}, {Csabai},
  {Czarapata}, {Davis}, {Doi}, {Dombeck}, {Eisenstein}, {Ellman}, {Elms},
  {Evans}, {Fan}, {Federwitz}, {Fiscelli}, {Friedman}, {Frieman}, {Fukugita},
  {Gillespie}, {Gunn}, {Gurbani}, {de Haas}, {Haldeman}, {Harris}, {Hayes},
  {Heckman}, {Hennessy}, {Hindsley}, {Holm}, {Holmgren}, {Huang}, {Hull},
  {Husby}, {Ichikawa}, {Ichikawa}, {Ivezi{\'c}}, {Kent}, {Kim}, {Kinney},
  {Klaene}, {Kleinman}, {Kleinman}, {Knapp}, {Korienek}, {Kron}, {Kunszt},
  {Lamb}, {Lee}, {Leger}, {Limmongkol}, {Lindenmeyer}, {Long}, {Loomis},
  {Loveday}, {Lucinio}, {Lupton}, {MacKinnon}, {Mannery}, {Mantsch}, {Margon},
  {McGehee}, {McKay}, {Meiksin}, {Merelli}, {Monet}, {Munn}, {Narayanan},
  {Nash}, {Neilsen}, {Neswold}, {Newberg}, {Nichol}, {Nicinski}, {Nonino},
  {Okada}, {Okamura}, {Ostriker}, {Owen}, {Pauls}, {Peoples}, {Peterson},
  {Petravick}, {Pier}, {Pope}, {Pordes}, {Prosapio}, {Rechenmacher}, {Quinn},
  {Richards}, {Richmond}, {Rivetta}, {Rockosi}, {Ruthmansdorfer}, {Sandford},
  {Schlegel}, {Schneider}, {Sekiguchi}, {Sergey}, {Shimasaku}, {Siegmund},
  {Smee}, {Smith}, {Snedden}, {Stone}, {Stoughton}, {Strauss}, {Stubbs},
  {SubbaRao}, {Szalay}, {Szapudi}, {Szokoly}, {Thakar}, {Tremonti}, {Tucker},
  {Uomoto}, {Vanden Berk}, {Vogeley}, {Waddell}, {Wang}, {Watanabe},
  {Weinberg}, {Yanny}, {Yasuda}, \& {SDSS Collaboration}}]{york_2000}
{York}, D.~G., {Adelman}, J., {Anderson}, John~E., J., {et~al.} 2000, \aj, 120,
  1579, \dodoi{10.1086/301513}

\end{thebibliography}
\bibliographystyle{aasjournal}



\end{document}